\documentclass[preprint,12pt]{elsarticle}
\usepackage[english]{babel}

\usepackage{natbib}

\usepackage{lineno,hyperref}
\usepackage{lscape}
\usepackage{xcolor}
\usepackage{color}
\usepackage{colortbl}
\usepackage{setspace}
\usepackage{amssymb}
\usepackage{amsmath}
\usepackage{calrsfs}
\usepackage{multirow}
\usepackage{dsfont}
\usepackage{stmaryrd}

\DeclareMathOperator*{\argmin}{arg\,min}
\DeclareMathAlphabet{\pazccal}{OMS}{zplm}{m}{n}

\modulolinenumbers[5]
\usepackage{dcolumn}
\usepackage{bm}
\usepackage{amsthm}
\usepackage[hang,flushmargin]{footmisc} 
\usepackage{graphics}
\usepackage{graphicx} 
\usepackage{mathrsfs}
\usepackage{bbm}
\usepackage{dsfont}
\usepackage{enumitem} 
\usepackage{subcaption}
\usepackage{booktabs}
\usepackage{sidecap}
\usepackage{xparse}
\usepackage{multirow}
\usepackage[margin=3cm]{geometry} 
\theoremstyle{definition}
\newtheorem{defn}{Definition}[section] 
\newtheorem{remark}{Remark}[section]
\newtheorem{prop}{Proposition}[section]

\setcitestyle{square}

\DeclareFontFamily{U}{cbgreek}{}
\DeclareFontShape{U}{cbgreek}{m}{n}{
        <-6>    grmn0500
        <6-7>   grmn0600
        <7-8>   grmn0700
        <8-9>   grmn0800
        <9-10>  grmn0900
        <10-12> grmn1000
        <12-17> grmn1200
        <17->   grmn1728
      }{}
\DeclareFontShape{U}{cbgreek}{bx}{n}{
        <-6>    grxn0500
        <6-7>   grxn0600
        <7-8>   grxn0700
        <8-9>   grxn0800
        <9-10>  grxn0900
        <10-12> grxn1000
        <12-17> grxn1200
        <17->   grxn1728
      }{}

\DeclareRobustCommand{\qoppa}{%
  \text{\usefont{U}{cbgreek}{\normalorbold}{n}\symbol{19}}%
}

\makeatletter
\newcommand{\normalorbold}{%
  \ifnum\pdf@strcmp{\math@version}{bold}=\z@ bx\else m\fi
}

\journal{---}
\begin{document}
\begin{frontmatter}



\title{Kolmogorov-Smirnov Estimation of Self-Similarity \\in Long-Range Dependent Fractional Processes}

\author[inst1]{Daniele Angelini}

\affiliation[inst1]{organization={MEMOTEF, Sapienza University of Rome},Department and Organization
            country={Italy}}

\author[inst1]{Sergio Bianchi}


\begin{abstract}
This paper investigates the estimation of the self-similarity parameter in fractional processes. We re-examine the Kolmogorov-Smirnov (KS) test as a distribution-based method for assessing self-similarity, emphasizing its robustness and independence from specific probability distributions. Despite these advantages, the KS test encounters significant challenges when applied to fractional processes, primarily due to intrinsic data dependencies that induce both intradependent and interdependent effects. 
To address these limitations, we propose a novel method based on random permutation theory, which effectively removes autocorrelations while preserving the self-similarity structure of the process. Simulation results validate the robustness of the proposed approach, demonstrating its effectiveness in providing reliable estimation in the presence of strong dependencies. These findings establish a statistically rigorous framework for self-similarity analysis in fractional processes, with potential applications across various scientific domains.
\end{abstract}


\end{frontmatter}


\section{Introduction} \label{sec:intro}
Many natural as well as artificial phenomena including turbulence, tree growth, gravitational waves or stock markets frequently exhibit complex patterns in which scaling or self-similarity emerges as a prominent characteristic \cite{CafieroVassilicos2019,KalliokoskiSievanenNygren2010,BaumgarteGundlachHilditch2023}. In these fields, self-similarity refers to the repetition of patterns or structures, and it plays a crucial role in uncovering meaningful features for applications, including pattern recognition, fractal analysis, and image compression. Its relevance spans a wide range of applications, including telecommunication networks \citep{Jeong2002}, hydrologic time series \citep{RaoBhattacharya1999}, and stock market prices \citep{ContPottersBouchad1997,Garcin2022}.

One of the parameters used to describe self-similarity in time series is the Hurst exponent $H\in\left(0,1\right]$. Introduced by Kolmogorov \citep{Kolmogorov1940}, it was related by Hurst to the phenomena of long memory \citep{Hurst1951,HurstBlackSimaika1965} and used by Mandelbrot and Van Ness to describe the non-markovian behaviour of the fractional Brownian motion (fBm), the unique mean-zero Gaussian process with stationary and self-similar increments \citep{MandelbrotVanNess1968}. FBm is the fractional extension of the standard Brownian motion: for $H>1/2$ the process is persistent and has long memory, while for $H<1/2$ the process exhibits antipersistence and is mean-reverting. Only when $H=1/2$ the fBm has independent increments and is a semi-martingale, falling into the standard Brownian motion \citep{AchardCoeurjolly2010}.

The property of self-similarity relates to the equivalence of the process's scaled distributions. Consequently, the Kolmogorov-Smirnov (KS) test \citep{Kolmogorov1950, Smirnov1939}, which assesses whether two univariate probability distributions are indistinguishable \citep{OleaPawlowsky-Glahn2009}, emerges as a natural choice to test whether a process exhibits self-similarity across different time scales. In contrast, the most popular estimators of the Hurst parameter—encompassing time-domain methods (e.g., R/S statistic \citep{MandelbrotTaqqu1979}, aggregated variance method \citep{Beran1994}, and Higuchi's method \citep{Higuchi1988}), frequency-domain methods (e.g., modified periodogram \citep{Hassler1993} and Whittle estimator \citep{TaqquTeverovsky1998}), and time-scale methods (e.g., Abry and Veitch's wavelet-based approach \citep{AbryVeitch1998} and the diffusion entropy method \citep{GrigoliniPalatellaRaffaelli2001})—rely on specific moments of the data (usually mean, absolute moments or variance) rather than analyzing how the entire distribution scales. This makes their robustness questionable to some extent, since they appear to introduce non-linear biases into the estimation process \citep{ContDas2022, ContDas2023, ContDas2024, AngeliniBianchi2023}.

In 2004 a new distribution-based method was proposed to detect the self-similarity parameter \citep{Bianchi2004}: the Hurst exponent is estimated as the value of $H$ that minimizes the KS statistic, evaluated with respect to the minimal and maximal elements of the space of rescaled probability distribution functions for a self-similar process. Being distribution-free, 
this method is robust with respect to infinite-moments self-similar stochastic processes. However, the KS statistic is designed for independent and identically distributed (i.i.d.) data. Consequently, significant challenges emerge when attempting to extend its application to dependent data:
\begin{itemize} [leftmargin=*]
    \item Fractional Brownian motion (fBm) or other fractional processes exhibit dependence when the $H \neq 1/2$, which complicates the estimation of the self-similarity parameter. For $H>1/2$, the presence of positive dependence among the data leads to an effective reduction in sample size, manifesting as an excessive rate of rejection of the null hypothesis for identical distributions (Type 1 error). Conversely, for $H<1/2$, the negative dependence results in an effective increase in sample size, leading to an excessive acceptance of the null hypothesis (Type 2 error). This poses a significant challenge in determining the critical threshold that governs the significance of the KS statistic, as its behavior exhibits a non-linear dependence on the Hurst exponent $H$. We term this issue the \textit{intradependent problem}.
    \item In estimating the self-similarity parameter, the KS test is performed between the rescaled probability distributions of the same starting sample. This fact introduces further dependence among the data when applying the KS test. In particular, it is observed from the results that the distribution 
    of the KS statistic depends not only on $H$ due to the intradependent problem, but also on the scaling factor of the distribution of a self-similar process. We call this effect as \textit{interdependent problem}.

\end{itemize}
The contribution of this paper is to overcome both the intradependent and interdependent problems by using random permutation theory to break the autocorrelation structure contained in fractional data without affecting their self-similarity structure. By circumventing the above problems, 
we provide a method which makes the KS statistic robust with respect to dependent data, even when the autocorrelation is very strong. Certainly, this is not the first attempt to make the KS statistic robust with respect to data dependence. To mitigate its influence, various strategies have been proposed in literature, but most of the solutions are either ad hoc or tailored to specific dependence models; as such, they are not entirely satisfactory. For example, \citep{Weiss1978} recommends a corrective approach for datasets generated by second-order auto-regressive processes with known parameters;
\citep{DurilleulLegendre1992} examine the effects of first-order autocorrelation and conclude that dependence leads to a reduction in the count of effectively independent observations, even if this bias diminishes to negligible levels when the first-order autocorrelation is within the range of $[-0.8,0.3]$ and as the sample size increases;
\citep{ChicheporticheBouchad2011} introduce a methodology that incorporates all lagged bivariate copulas to capture non-linear temporal dependencies, while \citep{Lanzante2021} evaluates three methodologies, determining that Monte Carlo simulations offer the most effective solution.\\
Investigating the impact of dependence on the critical values of the KS distribution is particularly relevant in the context of fBm. This is because the presence of long-term memory significantly influences the Effective Sample Size (ESS)\footnote{The ESS represents the equivalent number of independent observations in a dataset, accounting for autocorrelation. It adjusts for dependence, providing a more accurate measure of the dataset's informational content as if the observations were independent.}, as positive dependence reduces the statistical power of tests. This occurs because dependent observations contain less information than independent ones. The ESS from  $N$ observations with autocorrelation function $\rho$ is defined by \citep{ThiebauxZwiers1984} as
\begin{equation}\label{eq:ESS}
N_{eff} = N\left[\sum_{\tau=-(N-1)}^{N-1}\left(1-\frac{|\tau|}{N}\right)\rho(\tau)\right]^{-1}.
\end{equation}
From (\ref{eq:ESS}), it is easy to see that the ESS can exceed the number of observations in the sample in presence of negative autocorrelation ($H<1/2$), and this represents a critical vulnerability when applying the KS statistic to fractional processes. This occurs because successive observations provide more independent information compared to cases with or without positive autocorrelation. Several factors contribute to this phenomenon: (a) each observation introduces contrasting information relative to its predecessor, increasing the informational content per observation; (b) negative autocorrelation reduces estimator variance, akin to having a larger sample of independent observations; and (c) error terms offset each other more effectively than in purely random series, enhancing the statistical efficiency and reliability of parameter estimates.\\
Given these limitations, we propose a radically different approach based on disrupting the autocorrelation function without affecting the distributions being compared by the KS statistic.\\

The paper is organized as follows. In Section \ref{sec:method} we recall the method introduced in \citep{Bianchi2005}. 
Section \ref{sec:caseStudy} discusses the limits of applying the KS test to fBm due to interdependent and intradependent problems. 
In Section \ref{sec:random_method} random permutations are introduced to modify the initial method and solve these problems. Section \ref{sec:new_method} describes the steps of the new methodology, while -- through extensive simulations -- Section \ref{sec:simulation} provides evidence that the new methodology effectively corrects the biases of the KS test in estimating the self-similarity parameter of dependent data. Finally, Section \ref{sec:conclusion} concludes.

\section{Preliminaries and problem statament} \label{sec:method}

\subsection{Preliminaries}
This section will recall some basic definitions about the Hurst parameter estimation method introduced by \citep{Bianchi2004}.
\begin{defn}
    A nontrivial\footnote{The process $\{X_t\}_{t\geq0}$ is trivial if $X_t$ is a constant almost surely for avery $t$.} stochastic process $X=\{X_t\}_{t\geq0}$, stochastically continuous\footnote{The process $\{X_t\}_{t\geq0}$ is stochastically continuous at time $t$ if for any $\epsilon>0$ $\lim\limits_{h\to 0}\mathbb{P}\{\vert X_{t+h}-X_t\vert>\epsilon\}=0$} at $t=0$, is $self$-$similar$ with $H_0\geq0$ ($H_0$-$ss$), if for any $a\in\mathbb{R}^+$
    \begin{equation}\label{eq:H-ss}
        \{X_{at}\}_{t\geq 0} \overset{d}{=}\{a^{H_0}X_t\}_{t\geq 0}
    \end{equation}
    in the sense of finite-dimensional distributions\footnote{The finite-dimensional distributions of the stochastic process $\{X_t\}_{t\geq0}$ refer to the joint distributions of $\left(X_{t_1},\ldots,X_{t_m}\right)$ for fixed but arbitrary $0\leq t_1<t_2<\ldots<t_m$. The two stochastic processes $\{X_t\}_{t\geq0}$ and $\{Y_t\}_{t\geq0}$, defined on probability spaces $\left(\Omega,\pazccal{F},\mathbb{P}\right)$ and $(\tilde{\Omega},\tilde{\pazccal{F}},\tilde{\mathbb{P}})$ respectively and sharing the same state space $\left(\mathbb{R}^k,\pazccal{B}\left(\mathbb{R}^k\right)\right)$, are said to have the same finite-dimensional distributions if, for any integer $n\geq1$ and real numbers $0\leq t_1<t_2<\ldots<t_m$ and $A\in\pazccal{B}(\mathbb{R}^k)$, it is:\[\mathbb{P}((X_{t_1},\ldots,X_{t_m})\in A) = \tilde{\mathbb{P}}((Y_{t_1},\ldots,Y_{t_m})\in A).\]}.
\end{defn}
\begin{remark}\label{remark:1}
    If a $H_0$-$ss$ stochastic process $\{X_t\}_{t\geq0}$, with $X_0=0$ a.s., has stationary increments, the increment process is self-similar with the same parameter ($H_0$-$sssi$):
    \begin{equation}\label{eq:H-sssi}
        \{X_{t+a}-X_t\}_{t\geq0} \overset{d}{=} \{a^{H_0}(X_{t+1}-X_t)\}_{t\geq0}, \,\,\, \forall a\in\mathbb{R}^+.
    \end{equation}
    In fact, setting $t=1$ in (\ref{eq:H-ss})
    \[\{X_{a}\} \overset{d}{=} \{a^{H_0}X_1\},\]
    \[\{X_{a}-X_0\}\overset{d}{=}\{a^{H_0}(X_1-X_0)\},\]
    that is, exploiting the stationarity of the increments for any $t \geq 0$,
    \[ \{X_{t+a}-X_t\}\overset{d}{=}\{a^{H_0}\left(X_{t+1}-X_t\right)\}.\]
    that is equation \eqref{eq:H-sssi}.
\end{remark}
\begin{remark}
    If $\{X_t\}_{t\geq0}$ is nontrivial, stochastically continuous at $t=0$ and $H_0$-$ss$, $H_0=0$ if and only if $X_t = X_0$ almost surely for every $t>0$. If $\mathbb{E}\left[\vert X_1\vert^2\right]<\infty$, then $H_0\leq 1$ (see \citep{EmbrechtsMaejima2002}). To exclude degenerate cases, in the following we will assume that $0<H_0\leq 1$.
\end{remark}
Following the idea set forth in \citep{Bianchi2004}, it is possible to estimate the self-similarity exponent $H_0$ by testing the equality of the distributions in equation \eqref{eq:H-ss} (or equation \eqref{eq:H-sssi}). To this aim, the Kolmogorov-Smirnov test can be used.\\

Given a compact timescale set $\pazccal{A}=\left[\underline{a},\overline{a}\right]\subset\mathbb{R}^+$, denote by $\Phi_{X_{at}}(x)$  the cumulative distribution function of $X_{at}$, for any $a\in\pazccal{A}$. The equality in distribution in equation \eqref{eq:H-ss}, for a specific $H_0\in\left(0,1\right]$, can be written as:
\begin{equation}\label{eq:distr_H-ss1}
    \Phi_{X_{at}}(x) := \mathbb{P}\left(X_{at}<x\right) = \mathbb{P}\left(a^{H_0}X_t<x\right) = \Phi_{X_t}\left(a^{-{H_0}}x\right).
\end{equation}
Introducing the variable $H\in\left(0,1\right]$, equation \eqref{eq:distr_H-ss1} can be rewritten as
\begin{equation}\label{eq:distr_H-ss2}
    \Phi_{a^{-H}X_{at}}(x) := \mathbb{P}\left(a^{-H}X_{at}<x\right) = \mathbb{P}\left(a^{H_0-H}X_{t}<x\right) = \Phi_{X_t}\left(a^{H-H_0}x\right).
\end{equation}
Denoting as $\Psi_H := \{\Phi_{a^{-H}X_{at}}(x),a\in\pazccal{A},x\in\mathbb{R}\}$ the set of the distribution functions of $\{a^{-H}X_{at}\}_{t\geq0}$ and considering the distance function $\qoppa$ induced by the sup-norm $\vert\vert\cdot\vert\vert_{\infty}$ with respect to $\pazccal{A}$, the diameter of $\Psi_H$ on the metric space $\left(\Psi_H,\qoppa\right)$ is then defined as
\begin{align}
    \delta_{X_t}\left(\Psi_H\right) &:= \sup\limits_{x\in\mathbb{R}}\sup\limits_{a,b\in\pazccal{A}}\qoppa\left(x,a,b\right) \nonumber \\
    &= \sup\limits_{x\in\mathbb{R}}\sup\limits_{a,b\in\pazccal{A}}\vert \Phi_{a^{-H}X_{at}}(x)-\Phi_{b^{-H}X_{bt}}(x)\vert \nonumber \\
    &= \sup\limits_{x\in\mathbb{R}}\sup\limits_{a,b\in\pazccal{A}}\left\vert \Phi_{X_{t}}\left(a^{H-H_0}x\right)-\Phi_{X_{t}}\left(b^{H-H_0}x\right)\right\vert \nonumber \\
    &=\sup\limits_{x\in\mathbb{R}}\left\vert \Phi_{X_{t}}(\underline{a}^{H-H_0}x)-\Phi_{X_{t}}(\overline{a}^{H-H_0}x)\right\vert. \label{eq:distr_H-ss3}
\end{align}

In the case of stationary increments $Z_{t,a} := X_{t+a}-X_t$ it is possible to rewrite equations \eqref{eq:distr_H-ss1}, \eqref{eq:distr_H-ss2} and \eqref{eq:distr_H-ss3} respectively as
\begin{equation}\label{eq:distr_H-sssi1}
    \Phi_{Z_{t,a}}(x) := \mathbb{P}\left(Z_{t,a}<x\right) \overset{\text{using eq. \eqref{eq:H-sssi}}}{=} \mathbb{P}\left(a^{H_0}Z_{t,1}<x\right) = \Phi_{Z_{t,1}}\left(a^{-H_0}x\right),
\end{equation}
\begin{equation}\label{eq:distr_H-sssi2}
    \Phi_{a^{-H}Z_{t,a}}(x) := \mathbb{P}\left(a^{-H}Z_{t,a}<x\right) \overset{\text{using eq. \eqref{eq:H-sssi}}}{=} \mathbb{P}\left(a^{H_0-H}Z_{t,1}<x\right) = \Phi_{Z_{t,1}}\left(a^{H-H_0}x\right),
\end{equation}
and
\begin{equation}\label{eq:distr_H-sssi3}
    \delta_{Z_{t}}\left(\Psi_H\right) = \sup\limits_{x\in\mathbb{R}}\left\vert \Phi_{Z_{t,1}}\left(\underline{a}^{H-H_0}x\right)-\Phi_{Z_{t,1}}\left(\overline{a}^{H-H_0}x\right)\right\vert.
\end{equation}
The following propositions were proved for $\delta\left(\Psi_H\right)$ (for proofs see \citep{Bianchi2004}):
\begin{prop}\label{prop1}
    $\{X_t\}_{t\geq0}$ is $H_0$-$ss$ if and only if, for any $\pazccal{A}\subset\mathbb{R}^+$, $\delta_{X_t}\left(\Psi_{H_0}\right)=0$.
\end{prop}
\begin{prop}\label{prop2}
    Let $\{X_t\}_{t\geq0}$ be $H_0$-$ss$. Then $\delta_{X_t}\left(\Psi_H\right)$ is non-increasing for $H\leq H_0$ and non-decreasing for $H\geq H_0$.
\end{prop}
\begin{prop}\label{prop3}
    Let $\{X_t\}_{t\geq0}$, \textbf{x}$\geqq0$ or \textbf{x}$\leqq0$, $\{\pazccal{A}_i\}_{i=1,\ldots,n}$ be a sequence of timescale sets such that, denoted by $\underline{a}_i=\min\left(\pazccal{A}_i\right)$ and by $\overline{a}_i=\max\left(\pazccal{A}_i\right)$, it is $\underline{a}_i\leq\underline{a}_j$ and $\overline{a}_i\geq\overline{a}_j$ for $i>j$. Then, with respect to the sequence $\{\pazccal{A}_i\}$, $\delta_{X_t}\left(\Psi_H\right)$ is: non-decreasing, if $H\neq H_0$, or zero, if $H=H_0$.
\end{prop}
\begin{remark}
    Propositions \eqref{prop1}, \eqref{prop2} and \eqref{prop3} can be applied also to $\delta_{Z_{t}}\left(\Psi_H\right)$ in the case of $H$-$sssi$ processes. 
\end{remark}

\subsection{Problem statement}
The idea behind this approach is to estimate the self-similarity parameter $H_0$ by seeking the value of $H\in\left(0,1\right]$ that minimizes the Kolmogorov-Smirnov statistic $\delta_{X_{t}}\left(\Psi_H\right)$ of any pair of rescaled distributions of $X_t$. In fact, from Propositions \eqref{prop1} and \eqref{prop2}, the diameter $\delta_{X_t}\left(\Psi_H\right)$ of a self-similar process has a unique zero-valued minimum with respect $H\in\left(0,1\right]$, with the corresponding abscissa precisely equal to $H_0$. However, the zero-valued minimum is a merely theoretical concept, because from an empirical point of view one deals with a sample of length $N$ -- the observed path of the process $\{X_t\}$ -- from which one extracts two rescaled samples $X_{1,\cdot}$ and $X_{2,\cdot}$, of length $n$ and $m$ respectively. Thus, the empirical diameter
\begin{equation}\label{eq:empirical_delta}
    \hat{\delta}_{X_t}\left(\Psi_H\right) = \sup\limits_{x\in\mathbb{R}}\left\vert \Phi_{1,n}(x) - \Phi_{2,m}(x)\right\vert = \sup\limits_{x\in\mathbb{R}}\left\vert \frac{1}{n}\sum\limits_{i=1}^n\mathbbm{1}_{\underline{a}^{-H}X_{1,i}\leq x}-\frac{1}{m}\sum\limits_{i=1}^m\mathbbm{1}_{\overline{a}^{-H}X_{2,i}\leq x}\right\vert
\end{equation}
is always larger than zero. Therefore, the estimated self-similarity parameter is 
\begin{equation}\label{eq:empirical_H}
    \hat{H}_0 = \argmin_{H\in\left(0,1\right]}\,\,\hat{\delta}_{X_t}\left(\Psi_H\right)
\end{equation}
provided that $\hat{\delta}_{X_t}\left(\Psi_H\right)$ in equation (\ref{eq:empirical_delta}) can be considered negligible.\\

\begin{figure}[!ht]
    \centering
    \includegraphics[width=0.5\linewidth]{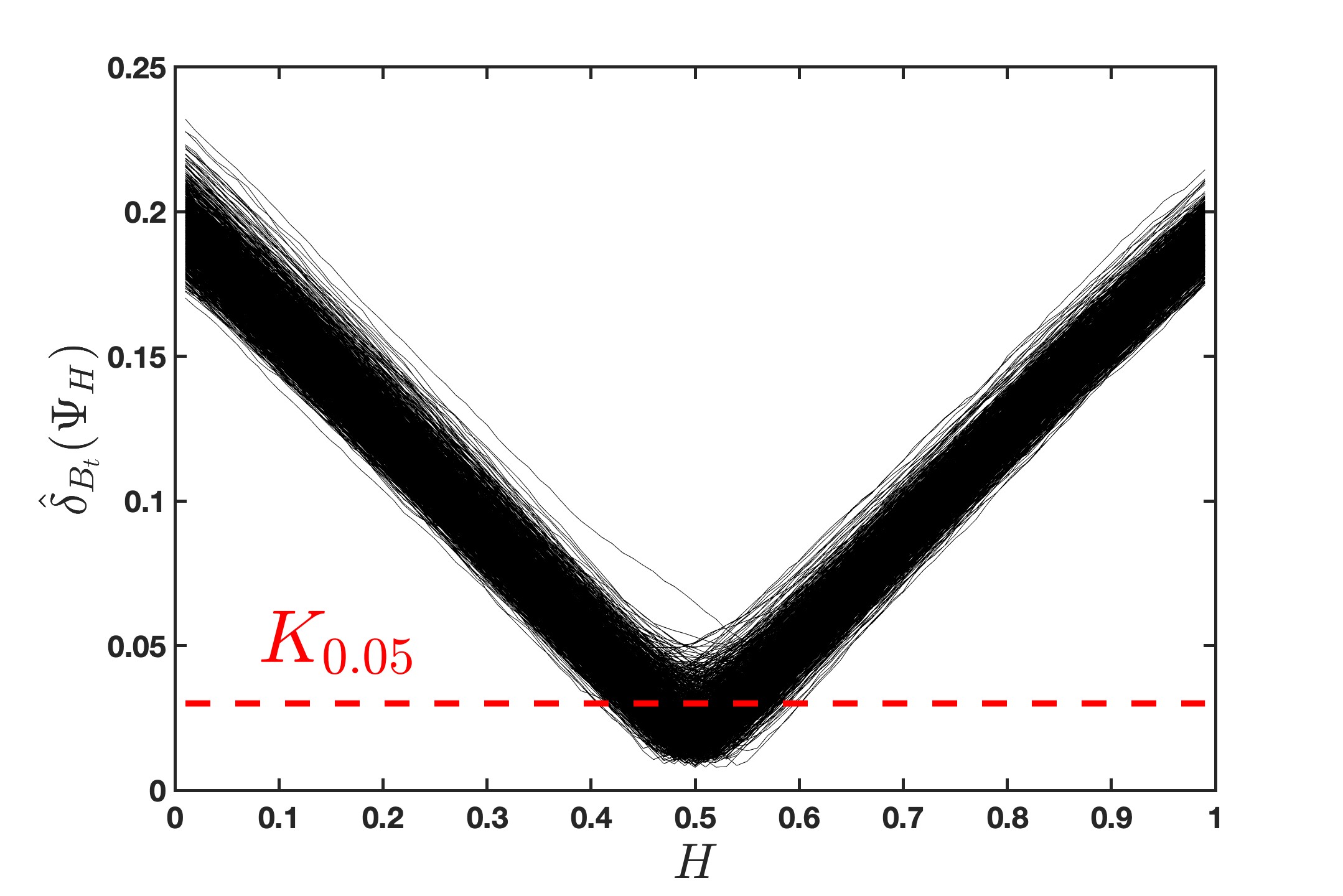}
    \caption{\footnotesize For any $H\in(0,1]$, with $\Delta H=0.01$, KS tests were performed between the CDFs of the rescaled increments of 1000 Brownian motion with $\underline{a}=1$ and $\overline{a}=5$. $\hat{H}_0$ is estimated as the minimum of the KS statistic in the $H$ domain. The mean value on all simulation is $\overline{\hat{H}}_0 = 0.4999\pm0.0006$. The red dotted line indicates the critical value for KS test at the significance level $\alpha=0.05$.}
    \label{fig:genericCase}
\end{figure}

It happens that the diameter $\hat{\delta}_{X_t}\left(\Psi_H\right)$ exactly corresponds to the Kolmogorov-Smirnov statistic $D_{n,m}$, which tests the null hypothesis that the two independent empirical cumulative distributions $\Phi_{1,n}$ and $\Phi_{2,m}$ are identical. For large samples, the null hypothesis $\Phi_{1,n} = \Phi_{2,m}$ is rejected at the significance level $\alpha$ if
\begin{equation}\label{eq:KS statistic}
    D_{n,m} = \sup\limits_{x}\left\vert \Phi_{1,n}(x)-\Phi_{2,m}(x)\right\vert > K_\alpha = \sqrt{-\log\left(\frac{\alpha}{2}\right)\frac{1+\frac{m}{n}}{2m}}.
\end{equation}
An application of this method is illustrated in Figure \ref{fig:genericCase}: 1000 sample paths of Brownian motion, each of length \( N = 4096 \), were generated. Two sets of increments, $\{B_{t+\underline{a}} - B_t\}$ and $\{B_{t+\overline{a}} - B_t\}$, were then extracted, with $\underline{a} = 1$ and $\overline{a} = 5$ , respectively\footnote{In \cite{Bianchi2004} it is proved that there is no loss of generality in setting $\underline{a}=1$.}. Since Brownian motion is self-similar with $H_0 = 1/2$, the diameter $\hat{\delta}_{B_t}\left(\Psi_H\right)$ passes the Kolmogorov-Smirnov test at a level $\alpha=0.05$ in the neighborhood of that value: $\overline{\hat{H}}_0 = 0.4999\pm 0.0006$.

\begin{figure}[h!]
  \footnotesize
\centering
  \includegraphics[width=250pt]{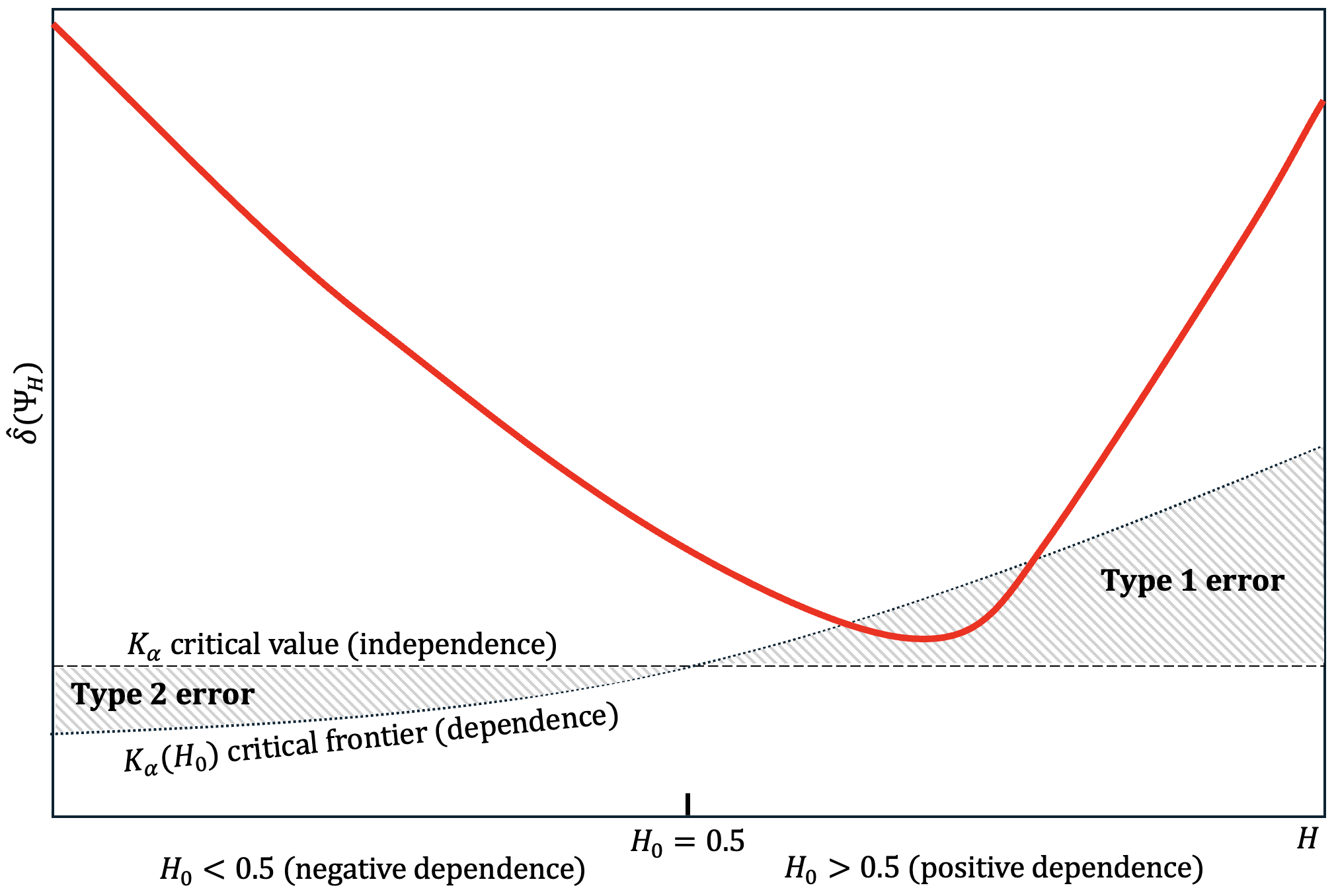}
  \caption{\label{fig:Sketched2} The critical value of the KS test \eqref{eq:KS statistic} refers to independence (dashed horizontal line), whereas dependence in the samples reflects in a new frontier of critical values (dotted line depending on $H$). The two frontiers intersect at the point $H=0.5$ (independence).}
  \normalsize
\end{figure}

The main challenge in applying this method to estimate the self-similarity parameter of fractional processes lies in their inherent dependence, which causes the distribution of the test statistic to deviate from \eqref{eq:KS statistic}, since this covers only the case of independent samples. For dependent fractional processes, a bias arises, leading to either a Type 1 or Type 2 error depending on the value of $H\neq1/2$ (see Figure \ref{fig:Sketched2}).
In general, when independence does not hold, the KS critical value tends to be too restrictive for moderate to high levels of positive autocorrelation and too lenient for moderate to high levels of negative autocorrelation. To mitigate the influence of dependence, various strategies have been proposed (as discussed in the Introduction, see e.g. \cite{Weiss1978,DurilleulLegendre1992,ChicheporticheBouchad2011,Lanzante2021}), but none appears to be satisfactory with even highly dependent data, as is the case of fractional processes with $H$ far from 0.5.

\section{KS distribution and fractional processes}\label{sec:caseStudy}
The method outlined in the previous Section can be applied to any self-similar process, provided of course that a solution is found to eliminate Type 1 and Type 2 errors from the significance analysis of statistic \eqref{eq:KS statistic}.
In the following, we will refer to the fractional Gaussian noise (fGn), the increment process of a fractional Brownian motion, as representative of the fractional (stationary) processes. 

Focusing solely on the case of fGn does not lead to a loss of generality, as the decorrelation technique introduced in Section \ref{sec:random_method} to remove the correlation structure of data while preserving their distributional properties, is applicable to nearly all fractional processes. This is due to the fact that, among fractional processes, fractional Gaussian noise (fGn) typically exhibits the slowest decay in its autocorrelation function (ACF), particularly for  $H > 0.5$, where the ACF follows a power-law decay of the form  $|\tau|^{2H-2}$. As shown in Table \ref{tab:acf-decay}, the autocorrelation functions of most fractional processes also decay as power laws, with a rate that is equal to or faster than that of fGn. Consequently, in the analysis presented in the subsequent sections, it suffices to focus on the fGn.  

\begin{table}[!ht]
\centering
\caption{\label{tab:acf-decay}Comparison of the speed of decay of autocorrelation functions for various Fractional Processes. fGn: Fractional Gaussian Noise; fPP: Fractional Poisson Process; fLm: Fractional Lévy Motion; fOU: Fractional Ornstein-Uhlenbeck (exponential decay); ARFIMA: Autoregressive Fractional Integrated Moving Average.}
\begin{tabular}{>{\raggedright}m{2.5cm} >{\centering}m{5cm} m{3.5cm}}
\toprule
\textbf{Process} &  \textbf{Asymptotic behaviour} & \textbf{Key Parameter} \\
\midrule
fGn & 
$\mathbb{E}[Z_{t}Z_{t+\tau}] \sim |\tau|^{2H-2}$ & 
 \( H \in (0,1) \) \\
\vspace{.2cm}
fPP &  
$\mathbb{E}[N(t)N(t+\tau)] \sim |\tau|^{-\alpha}$ & 
 \( \alpha \in (0,1) \) \\
\vspace{.2cm}
fLm & 
$\mathbb{E}[L_H(t)L_H(t+\tau)] \sim |\tau|^{2H-2}$ & 
 \( H \in (0,1) \) \\
\vspace{.2cm}
fOU & 
$\mathbb{E}[X(t)X(t+\tau)] \sim |\tau|^{2H-2}$ & 
 \( H \in (0,1) \) \\
\vspace{.2cm}
ARFIMA & 
$\rho(\tau) \sim |\tau|^{2d-1}$ & 
 \( d \in (0, 0.5) \) \\
\bottomrule
\end{tabular}
\vspace{.2cm}
\end{table}

\subsection{Fractional Brownian motion}
For the reasons discussed in the previous paragraph, it is worthwhile to recall some properties of the fBm. A fractional Brownian motion (fBm) with Hurst exponent $H\in\left(0,1\right]$ is a centered Gaussian, non-stationary and self-similar process $\{B^{H}_t\}_{t\geq 0}$ defined as the following integral moving average representation \cite{Coeurjolly2000} with respect to the Brownian motion $B_t$ 
\begin{equation}\label{eq:fBm_MovingAverage}
    B_{t}^{H} = \frac{CV_H^{1/2}}{\Gamma\left(H+\frac{1}{2}\right)}\int_{\mathbb{R}}\left[(t-s)_+^{H-\frac{1}{2}} - (-s)_+^{H-\frac{1}{2}}\right]dB_s,
\end{equation}
where $C$ is a scale parameter, $V_H = \Gamma(2H+1)\sin\left(\pi H\right)$, $\Gamma$ denotes the Gamma function and $x_+ = \max\left(x,0\right)$. From the covariance function
\begin{equation}
K_{B^H}(t,s)=\mathbb{E}\left(B_t^{H}B_s^{H}\right)=\frac{C^2}{2}\left(\vert t\vert^{2H}+\vert s\vert^{2H}-\vert t-s\vert^{2H}\right),
\end{equation}
it is clear that the scaling parameter $C$ is introduced such that the variance at unit time is equal to $C^2$. The peculiarity of the fBm is that, up to a multiplicative constant, it is the unique mean-zero Gaussian process with stationary and self-similar increments.

Regarding the stationarity of the increments $Z_{t,a}^H = B_{t+a}^{H} - B_{t}^{H}$, for any positive increment $a > 0$, we recall that $\mathbb{E}\left(Z_{t,a}^H\right) = 0$ and that the covariance function is independent of time:
\begin{align} \label{eq:ACFfGn}
K_{Z_{\cdot,a}^H}(t-s) &= \mathbb{E}\left(Z_{t,a}^HZ_{s,a}^H\right) \nonumber \\
& = \mathbb{E}\left[(B_{t+a}^H-B_t^H)(B_{s+a}^H-B_s^H)\right] \nonumber \\
&= \mathbb{E}(B_{t+a}^HB_{s+a}^H)-\mathbb{E}(B_{t+a}^HB_s^H)-\mathbb{E}(B_t^HB_{s+a}^H)+\mathbb{E}(B_t^HB_s^H) \nonumber \\
 &= \frac{C^2}{2}\left(\vert t+a-s\vert^{2H}+\vert t-a-s\vert^{2H}-2\vert t-s\vert^{2H}\right), \; t,s\geq0.
\end{align}

\begin{figure}[!ht]
\centering
    \begin{subfigure}{1\textwidth}
\centering
        \includegraphics[width=12cm,height=5cm]{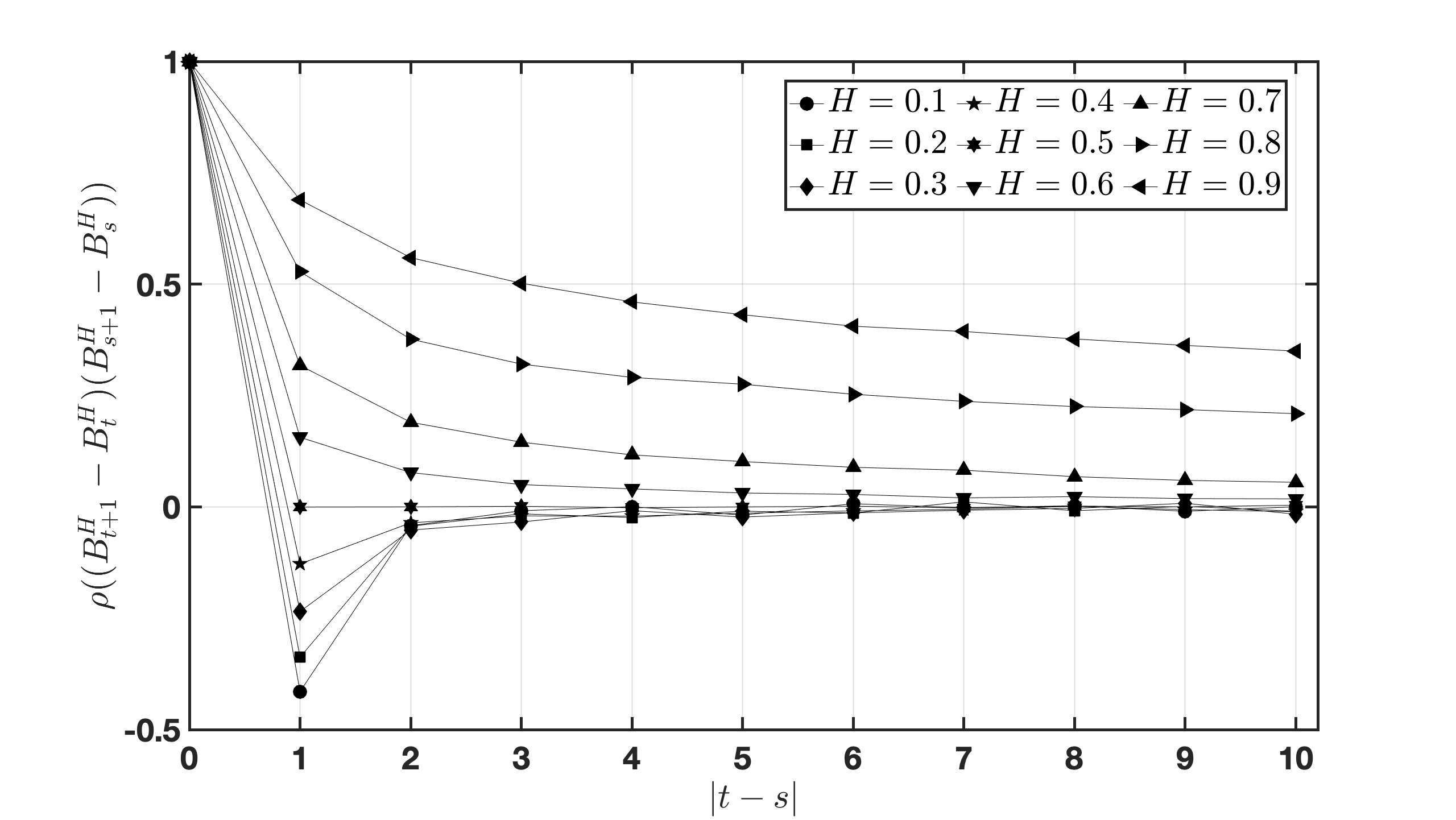}
        \caption{}
        \label{fig:ACF_incr_fbm}
    \end{subfigure}
    \begin{subfigure}{0.42\textwidth}
        \centering
        \includegraphics[width=5cm,height=5cm]{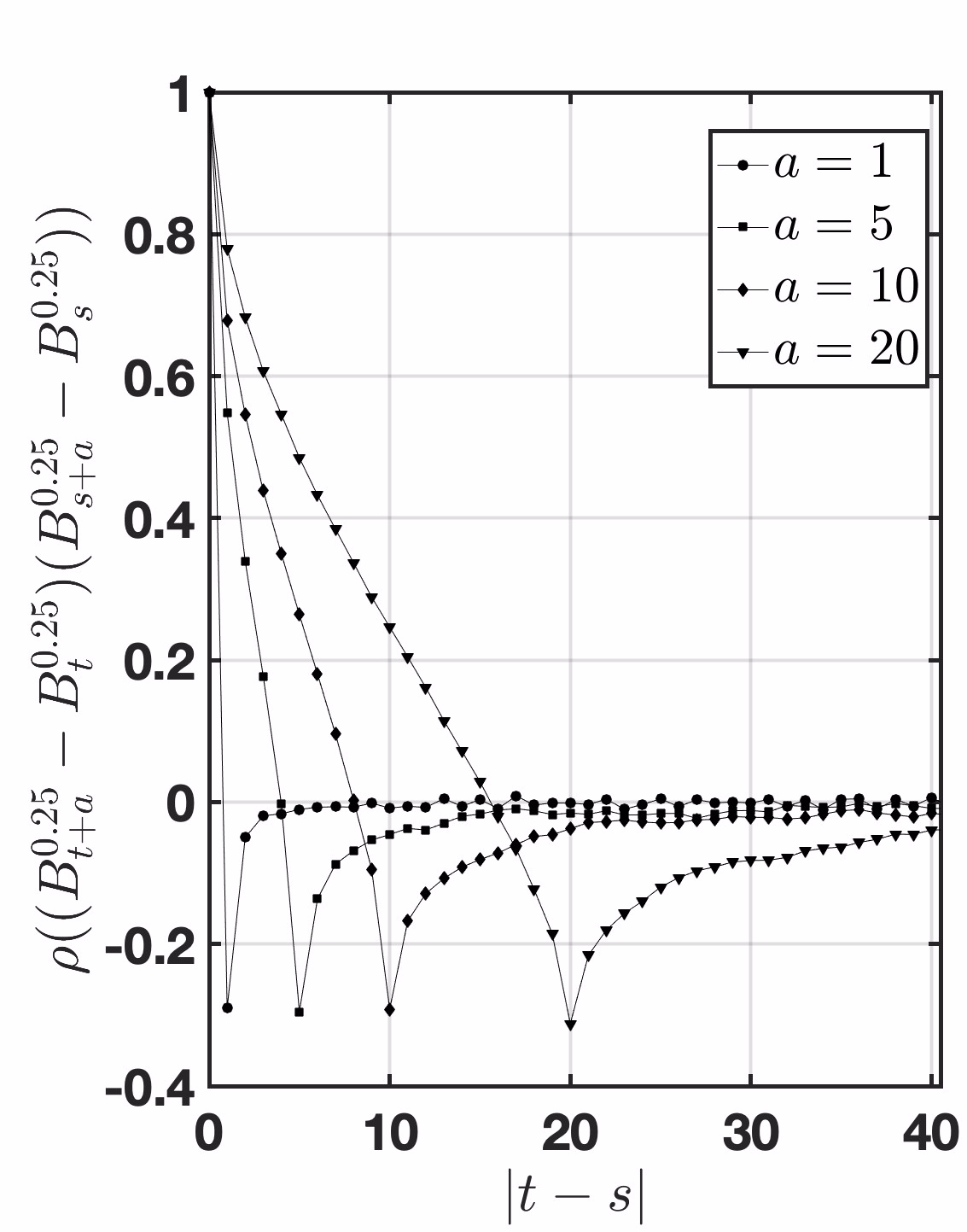}
        \caption{}
        \label{fig:ACF_NoMixed_min}
    \end{subfigure}
    \hspace{-1.3cm}
    \begin{subfigure}{0.42\textwidth}
        \centering
        \includegraphics[width=5cm,height=5cm]{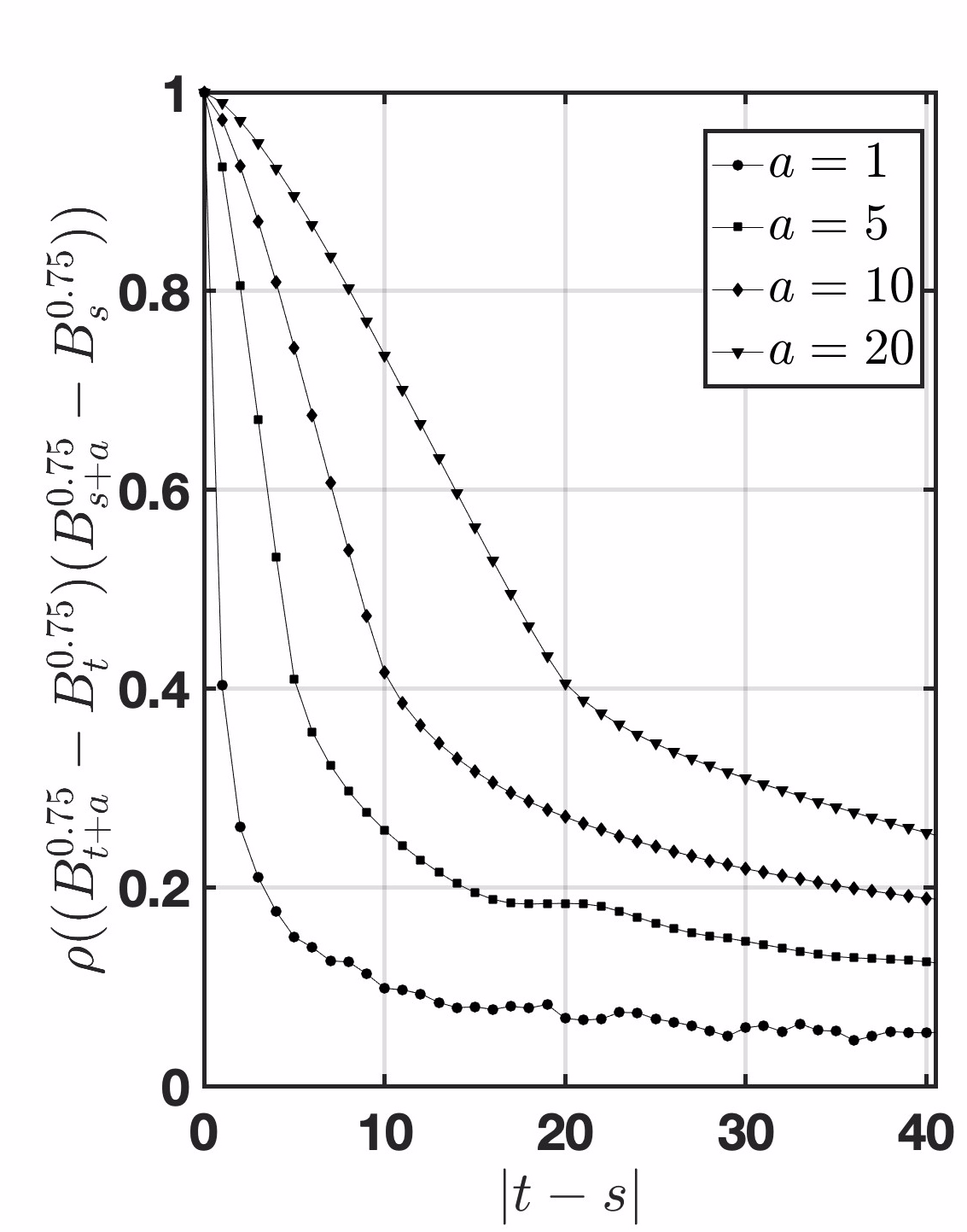}
        \caption{}
        \label{fig:ACF_NoMixed_max}
    \end{subfigure}
    \caption{\footnotesize Autocorrelation function (ACF) $\rho$ for the increments of a fractional Gaussian noise $B_{t+a}^H-B_{t}^H$. In this figure it was studied: (a) the ACF for the unit increment for many values of the Hurst exponent $H$; the ACF for such time scale parameters $a=\{1,5,10,20\}$ (b) in the antipersistent case $H=0.25$ and (c) in the persistent case $H=0.75$.}
    \label{fig:ACF1}
\end{figure}
Therefore, as shown in Remark (\ref{remark:1}), the non-markovian behaviour of the fBm also affects the increment process, called fractional Gaussian noise (fGn). In Figure \ref{fig:ACF1} it is represented the autocorrelation function $\rho\left(Z_{t,a}^{H}Z_{s,a}^{H}\right)$ which depends only on the temporal lag $\vert t-s\vert$ and the increments $a$. Panel (a) illustrates the effect of the Hurst exponent $H$ on memory: for $H > 0.5$, the increments exhibit long memory, whereas for $H < 0.5$, they display short memory. In panel (c), corresponding to $H > 0.5$, we observe that the autocorrelation increases significantly as the increment $a$ grows, compared to the short-memory case shown in panel (b).

\subsection{Analysis of the bias for the fGn}
Applying equations \eqref{eq:distr_H-sssi1}-\eqref{eq:distr_H-sssi3} to an fBm with Hurst exponent $H_0$, we have
\begin{equation}
    \Phi_{Z_{t,a}^{H_0}}(x) :=\mathbb{P}\left(Z_{t,a}^{H_0}<x\right)=\mathbb{P}\left(a^{H_0-H}Z_{t,1}<x\right) = \Phi_{Z_{t,1}^{H_0}}\left(a^{H_0-H}x\right), \;\;\;\text{for any }a>0. 
\end{equation}
Given a timescale set $\pazccal{A}=\left[\underline{a},\overline{a}\right]$, with $\underline{a}=1$, the diameter becomes
\begin{equation}
\delta_{Z^{H_0}_{t}}(\Psi_H)=\sup\limits_{x\in\mathbb{R}}\left\vert \Phi_{Z_{t,1}^{H_0}}(x)-\Phi_{Z_{t,1}^{H_0}}\left(\overline{a}^{H-H_0}x\right)\right\vert.
\end{equation}
\begin{figure}[!ht]
    \centering  \includegraphics[width=0.5\linewidth]{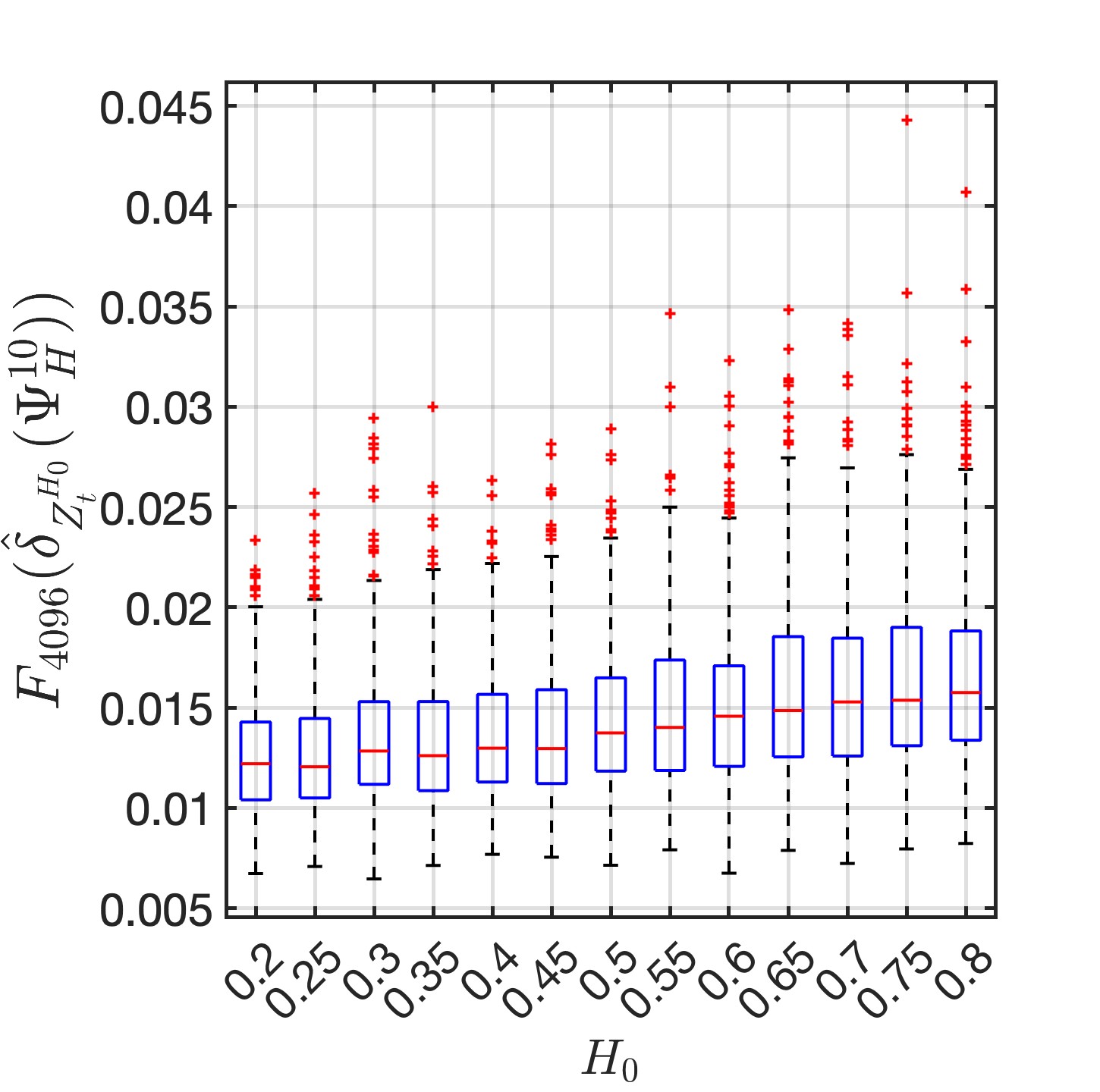}
    \caption{\footnotesize Distributions $F_N$ of the minimum of the empirical KS statistics $\hat{\delta}$ performed between rescaled distribution of fractional Gaussian noise $Z_{t,1}^{H_0}$ and $Z_{t,\overline{a}}^{H_0}$ for $H_0\in[0.2,0.8]$ with $\Delta H_0=0.05$. This experiment reveals that the distributions are influenced by the parameter $H_0$, as the Hurst exponent introduces a positive or negative correlation depending on its value. This violates one of the assumptions of the KS test, i.e. independence between the series data. In this experiment it is used $N = 4096$, $\overline{a}=10$ and it is performed $M=500$ simulations for each $H_0$.}
    \label{fig:delta VS H no mixed}
\end{figure}
In a computational analysis we simulate\footnote{The fractional Brownian motion was generated using the \textit{fbmwoodchan()} MATLAB function, available in the FracLab Toolbox 2.02 (INRIA package), which is based on the Wood and Chan circulant matrix method \citep{WoodChan1994}. Wood and Chan's method simulates very accurate fGn, but this simulator returns fBm's as a cumulative sum of fGn's. As shown in \citep{Coeurjolly2000}, this approximation results in errors in the covariance structure of fBm's generated as $H$ increases. For this reason the simulations in this paper will stop at the value of $H=0.8$. By symmetry, the domain considered is $H\in[0.2,0.8]$.} an fBm $\left\{B_{i}^{H_0}\right\}_{i\in\llbracket 0,N\rrbracket}$ with $t=\frac{i}{N}$. From the same process we extract two dependent increment samples $Z_{i,\underline{a}}^{H_0} =Z_{i,1}^{H_0} = B_{i+1}^{H_0}-B_i^{H_0}$ and $Z_{j,\overline{a}}^{H_0} = B_{j+\overline{a}}^{H_0}-B_j^{H_0}$, with $i\in\llbracket 0,N-1\rrbracket$ and $j\in\llbracket 0,N-\overline{a}\rrbracket$ respectively. Therefore, setting $\Psi_H^{\overline{a}} := \{\Phi_{a^{-H}Z_{t,a}}(x),a\in \{1,\overline{a}\},x\in\mathbb{R}\}$, the empirical diameter is
\begin{equation}
    \hat{\delta}_{Z_t^{H_0}}(\Psi_H^{\overline{a}}) = \sup\limits_{x\in\mathbb{R}}\left\vert \frac{1}{N}\sum\limits_{i=0}^{N-1}\mathbbm{1}_{Z_{i,1}^{H_0}\leq x}-\frac{1}{N-\overline{a}+1}\sum\limits_{i=0}^{N-\overline{a}}\mathbbm{1}_{\overline{a}^{-H}Z_{i,\overline{a}}^{H_0}\leq x} \right\vert.
\end{equation}
The estimated Hurst exponent is
\begin{equation}
    \hat{H}_{H_0}^{\overline{a}} = \argmin_{H\in\left(0,1\right]} \,\,  \hat{\delta}_{Z_{t}^{H_0}}(\Psi_H^{\overline{a}}).
\end{equation}
Given $M = 500$ simulations with the same parameters $H_0$, $N=4096$, $\underline{a}=1$ and $\overline{a}=10$, we consider the distribution of the empirical minimum diameter $F_N(\hat{\delta}_{Z_t^{H_0}}\left(\Psi^{\overline{a}}_{H}\right))$ and the distribution of the estimated Hurst exponent $G_N\left(\hat{H}^{\overline{a}}_{H_0}\right)$. In Figure \ref{fig:delta VS H no mixed}, $F_N$ is shown for $H_0\in\left[0.2,0.8\right]$ with $\Delta H_0 = 0.05$. This experiment confirms the bias discussed in Figure \ref{fig:Sketched2} and shows that the distribution of the KS statistic increases as $H_0$ increases, precisely because the test is sensitive to autocorrelation: in the case of positive correlation ($H_0 > 0.5$), the effective number of independent data points is artificially reduced, leading to an increase in the critical value $K_{\alpha}$. Conversely, when there is a negative correlation ($H_0 < 0.5$), the effective number of independent data points is artificially inflated, resulting in a decrease in the critical value $K_{\alpha}$. We refer to this problem as the \textit{intradependent problem}.
\begin{figure}[!ht]
    \centering  \includegraphics[width=1\linewidth]{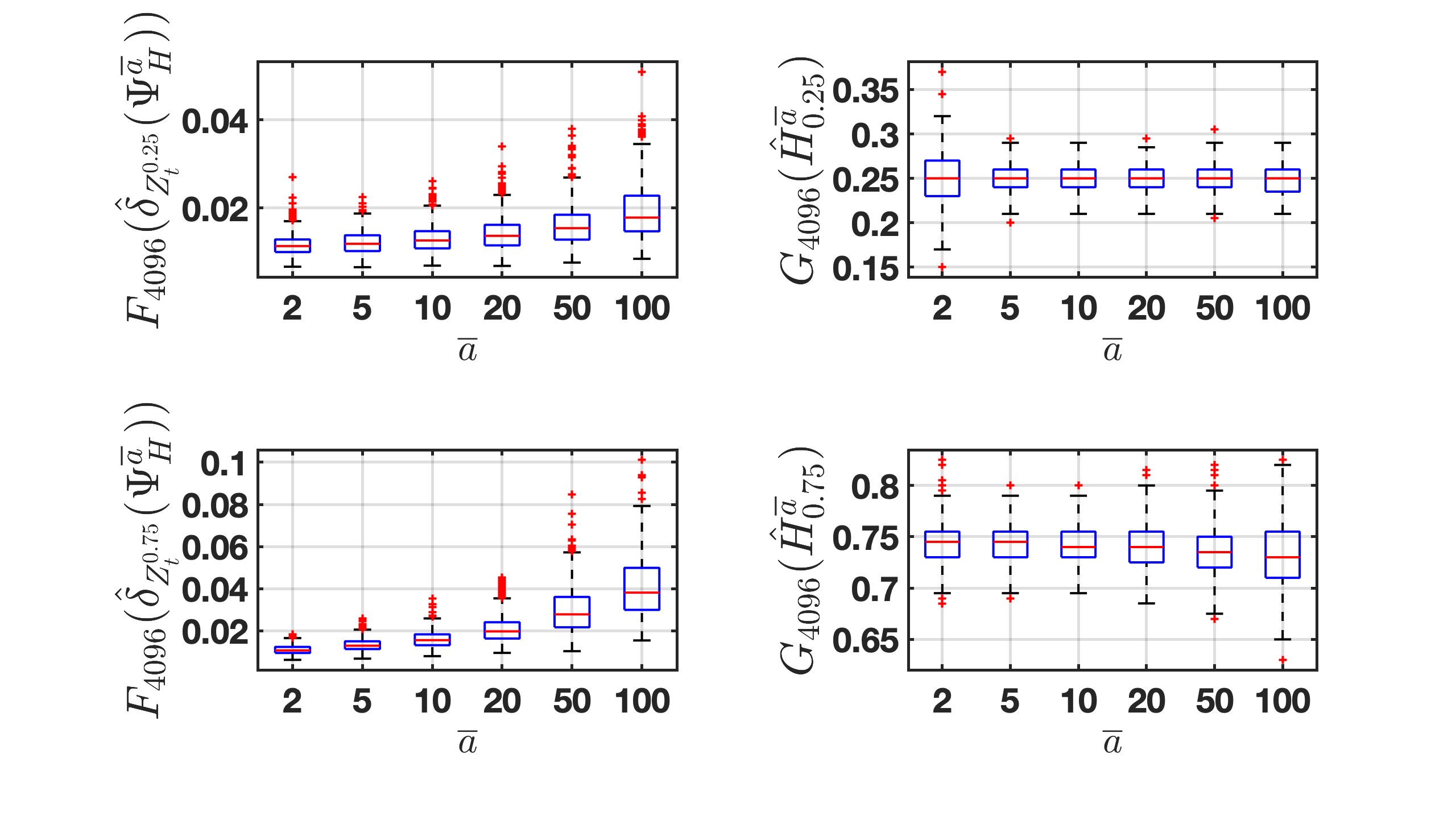}
    \caption{\footnotesize Distributions $F_N$ and $G_N$, performed between the rescaled distributions of fractional Gaussian noise $Z_{t,1}^{H_0}$ and $Z_{t,\overline{a}}^{H_0}$, of the minimum of the empirical KS statistics $\hat{\delta}$ and the estimated Hurst exponent. For each $H_0$, the $M=500$ simulations were performed with $N = 4096$. In upper panels the experiment was run with $H_0 = 0.25$ and in lower panels with $H_0=0.75$: in each case the distributions $F_N$ increase as value of $\overline{a}$ increases. The experiment is in contrast with the theoretical assumption that the distribution of $\delta$ does not depend by $\overline{a}$.}
    \label{fig:delta VS a no mixed}
\end{figure}
In Figure \ref{fig:delta VS a no mixed} we carried out two experiments, setting $H_0$ equal to 0.25 and 0.75, where we studied the distribution of the empirical diameters $F_N$, and of the respective distributions of the estimated Hurst exponent $G_N$, varying $\overline{a}={2,5,10,20,50,100}$. 
According to the method outlined in Section \ref{sec:method}, and in particular the results provided by Propositions \eqref{prop1}, \eqref{prop2}, and \eqref{prop3}, the distribution of diameters \( \hat{\delta} \) should be independent of the chosen value of \( a \). However, in Figure \ref{fig:delta VS a no mixed}, we observe that the distribution \( F_N \) increases as \( \overline{a} \) grows. This discrepancy between theory and empirical results arises from the fact that the method involves extracting two rescaled series from the same original series, leading to a high correlation between the two series on which the KS test is applied. We refer to this issue as the \textit{interdependent problem}.

\section{Decorrelation by random permutation}\label{sec:random_method}
The intradependent issue stems from the fact that the fGn exhibits the autocorrelation function given by \eqref{eq:ACFfGn}, whose effects are illustrated in Figure \ref{fig:ACF1}. To overcome the intradependent problem, it is necessary to eliminate the dependence between increments and, consequently, disrupt the autocorrelation function of the fGn. To achieve this and to ease the understanding of the next proposition, it is useful to recall some fundamental concepts regarding the power spectrum and random permutation theory, since both tools will be used in the sequel.\\

Since the power spectrum $S_{X}(\omega)$ of a stationary process $\{X_n\}_{n\in\mathbb{Z}}$ is defined as the Fourier transform of its covariance function, one has that the autocovariance function can be obtained as the inverse Fourier transform of its power spectrum, i.e.
    \begin{equation}
        K_{X}(q)=\mathbb{E}[X_nX_{n-q}] = \int_{-\pi}^{\pi}e^{iq\omega}S_X(\omega)d\omega.
    \end{equation}
    The power spectrum of the fractional Gaussian noise $Z^H_{t,a}$, for any $a>0$,  is defined as
    \begin{equation}\label{eq:powerfgn}
        S_{Z^H}(\omega) = 2 c_H(1-\cos\omega)\sum_{j\in\mathbb{Z}}\vert2\pi j + \omega\vert^{-1-2H}, \;\;\;\;\;\forall\omega\in[-\pi,\pi],
    \end{equation}
    with $c_H = \frac{C^2}{2\pi}\sin(\pi H)\Gamma(2H+1)$ \citep{Coeurjolly2000}.
To disrupt the structure of   \eqref{eq:powerfgn}, we need to introduce the random permutations of a sequence. Given the random sequence $Z = \{Z_\ell, \ell\in\mathbb{Z}\}$, we create a new sequence $U = \{U_\ell, \ell\in\mathbb{Z}\}$ such that 
 \begin{equation}\label{eq: new sequence}
     U_\ell = Z_{\overline{\ell}L+b_{\underline{\ell}}}, \;\; \ell = \overline{\ell}L+\underline{\ell},\;\; \overline{\ell}\in\mathbb{Z},\;\; 0\leq\underline{\ell}\leq L-1,
 \end{equation}
 where $b = (b_0, b_1, \ldots, b_{L-1})$ is a random permutation of the vector $(0,1,\ldots,L-1)$, uniformly distributed such that $\mathbb{P}(b) = \frac{1}{L!}$, with $\overline{\ell}$ and $\underline{\ell}$ the quotient and the remainder, respectively, of $\frac{\ell}{L}$.
Based on Section 4 in \citep{LacazeRoviras2002}, we can formulate the following \\

\begin{figure}[!ht]
    \centering
    \includegraphics[width=1\linewidth]{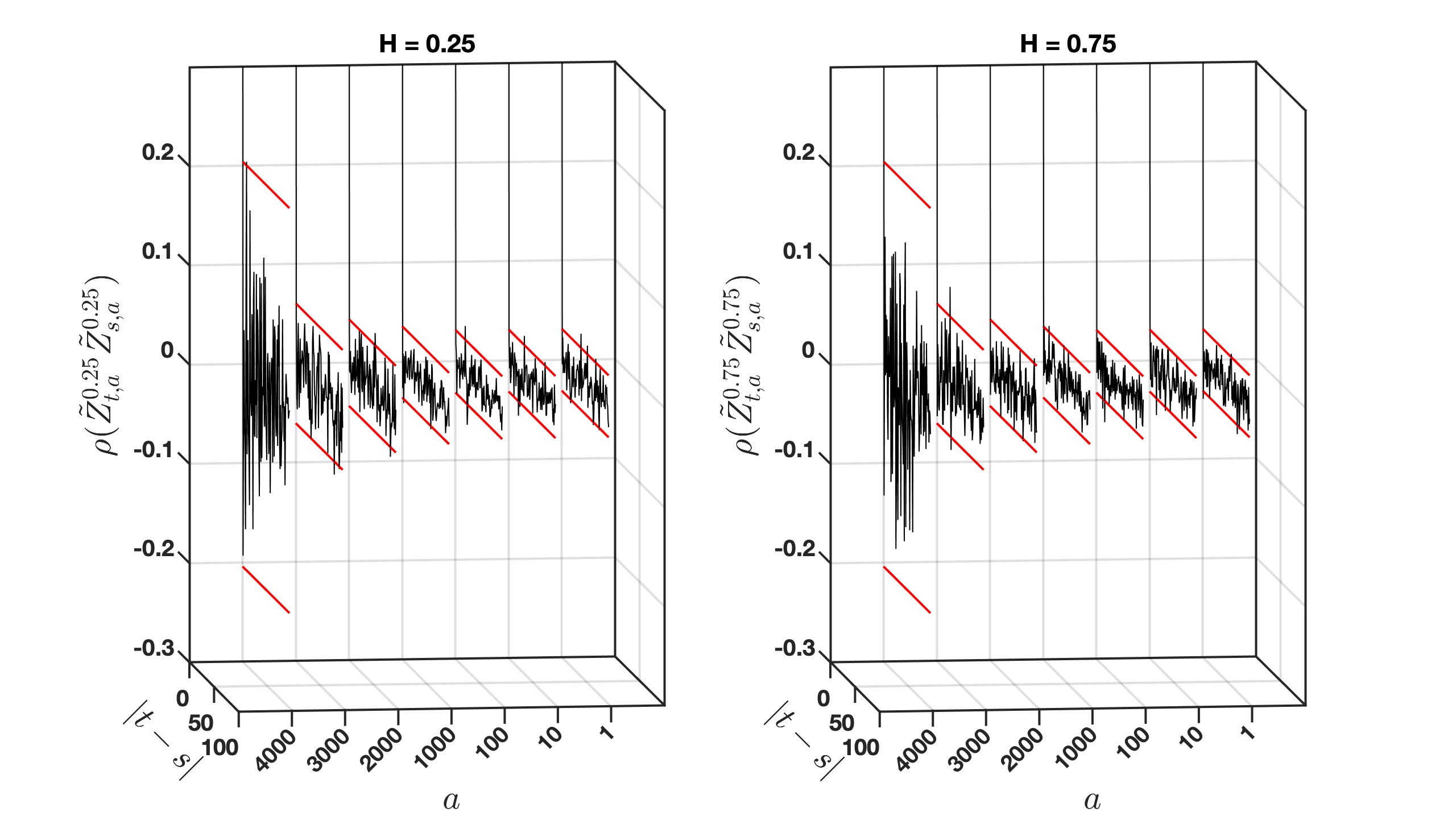}
    \caption{Autocorrelation function of the $\tilde{Z}^H_{t,a}$, the result of the random permutation of an fGn $Z^H_{t,a}$, for different value of scaling $a$ and $N=4096$. For any $a$ the new process $\tilde{Z}^H_{t,a}$ behaves like a white noise, also for $a$ near to the length of the process. The red lines represent the upper and lower $95\%$-confidence bounds assuming the input series is white noise.}
    \label{fig:ACF_mixed}
\end{figure}

 \begin{prop}\label{prop:1}
     Let $Z^H = \{Z^H_{\ell,a},\ell\in\mathbb{Z},a>0\}$ be a fractional Gaussian noise. Then, the new sequence $\tilde{Z}^H = \{Z^H_{\overline{\ell}L + b_{\underline{l}}+\phi,a},\ell\in\mathbb{Z},a>0\}$, with an independent random phase $\phi$ uniformly distributed on the set of integers $(0,1,\ldots,L-1)$, has factorizable covariance in the limit of large permutation length $L$
     \begin{equation}
         \lim\limits_{L\to\infty}K_{\tilde{Z}^H}(q) = \mathbb{E}\left[Z^H_{\overline{\ell}L + b_{\underline{l}}+\phi,a}\right]\mathbb{E}\left[Z^H_{\overline{\ell+q}L + b_{\underline{l+q}}+\phi,a}\right] = 0
     \end{equation}
     for $q\neq 0$.
 \end{prop}
 \begin{proof}
 As in equation \eqref{eq: new sequence}, under a random permutation $b=(b_0,b_1,\ldots,b_{L-1})$, we can rewrite the fGn $Z^H$ as the new sequence $U^H$, whose covariance function is 
     \begin{equation}
         K_{U^H}(\ell,\ell-q) = \mathbb{E}\left[Z^H_{\overline{\ell}L+b_{\underline{\ell}}}Z^H_{\overline{\ell-q}L+b_{\underline{\ell-q}}}\right] = \int_{-\pi}^{\pi}e^{i\left(\overline{\ell}-\overline{\ell-q}\right)L\omega}\mathbb{E}\left[e^{i\left(b_{\underline{\ell}}-b_{\underline{\ell-q}}\right)\omega}\right]S_{Z^H}(\omega)d\omega.
     \end{equation}
     Using the equality in distribution $\{b_{\underline{\ell}}-b_{\underline{\ell-q}}\} \overset{d}{=}\{b_1 - b_2\}$ for any $\underline{q}\neq 0$
     \begin{align}
    d_L(\omega) = \mathbb{E}\left[e^{i\left(b_{\underline{\ell}}-b_{\underline{\ell-q}}\right)\omega}\right] &= \mathbb{E}\left[e^{i\left(b_1-b_2\right)\omega}\right]\nonumber \\
    &=\frac{1}{L(L-1)}\sum\limits_{j,k=0,j\neq k}^{L-1}e^{i(j-k)\omega} \nonumber\\
    &=\frac{1}{L(L-1)}\left(\left\vert\frac{1-e^{-iL\omega}}{1-e^{-i\omega}}\right\vert^2-L\right)\nonumber\\
    &=\frac{1}{L(L-1)}\left(\left(\frac{\sin(L\omega/2)}{\sin(\omega/2)}\right)^2 - L\right).
\end{align}
     In short, for any $\underline{q}$, we have
     \begin{equation}
         \mathbb{E}\left[e^{i\left(b_{\underline{\ell}}-b_{\underline{\ell-q}}\right)\omega}\right] = 
         \begin{cases}
            1,\,\;\;\;\;\;\;\;\; \underline{q} = 0\\
             d_L(\omega),\;\; \underline{q}\neq 0
         \end{cases}.
     \end{equation}
     $U^H$ is not generally stationary, but it becomes stationary by adding an independent random phase $\phi$ uniformly distributed on the set of integers $\{0,1,\ldots,L-1\}$ with probability $\frac{1}{L}$. We obtain $\tilde{Z}^H=\{\tilde{Z}^H_\ell,\ell\in\mathbb{Z}\}$, such that $\tilde{Z}_\ell^H = U^H_{\ell+\phi}$, with covariance function
     \begin{equation}
         K_{\tilde{Z}^H}(\ell - (\ell-q)) = K_{\tilde{Z}^H}(q) =\int_{-\pi}^{\pi}e^{iq\omega}S_{\tilde{Z}^H}(\omega)d\omega.
     \end{equation}
     As shown in Appendix A of \citep{LacazeRoviras2002}, $\tilde{Z}^H$ is stationary and depends on both the initial power spectrum $S_{Z^H}(\omega)$ and the disorder of the random permutation $b$. The spectral density $S_{\tilde{Z}^H}(\omega)$ is then defined by
     \begin{equation}
         K_{\tilde{Z}^H}(q) = 
         \begin{cases}
             K_{Z^H}(q), \;\;\;\;\;\;\;\;\;\;\;\;\;\;\;\;\;\;\;\;\;\;\;\;\;\;\;\;\;\;\;\;\;\;\;\;\;\;\;\;\;\;\;\;\;\;\;\;\;\;\;\;\;\;\;\;\; \underline{q} =0\\
             \frac{1}{L}\int_{-\pi}^\pi e^{i\overline{q}L\omega}\left[\underline{q}e^{iL\omega}+(L-\underline{q})\right]d_L(\omega)S_{Z^H}(\omega)d\omega,\;\underline{q}\neq0
         \end{cases}.
     \end{equation}
     Noting that \textit{i}) $\vert d_L(\omega)\vert\leq1$, \textit{ii})$\lim\limits_{L\to\infty}d_L(\omega) = 0$ uniformly in $\omega$ in any interval that does not contain $\omega = 0$, and \textit{iii}) $\lim\limits_{L\to\infty}\ell = \lim\limits_{L\to\infty}(\overline{\ell}L+\underline{\ell}) = \underline{\ell}$, we have     
     \begin{equation}\label{eq: cov white noise}
         \lim\limits_{L\to\infty}K_{\tilde{Z}^H}(q) = \lim\limits_{L\to\infty}\int_{-\pi}^{\pi}e^{iq\omega}S_{\tilde{Z}^H}(\omega)d\omega =
         \begin{cases}
             K_{Z^H}(\omega),\, q=0\\
             0,\;\;\;\;\;\;\;\;\;\;\; q\neq0
         \end{cases}.
     \end{equation}
     The proof is concluded noting that equation \eqref{eq: cov white noise} is the covariance function of a white noise.
 \end{proof}
 \begin{remark}
    Given the stationarity of a fractional Gaussian noise, aside from a multiplicative parameter, we can state that $S_{Z^H}(\omega)$ is the power spectrum of its autocorrelation function.
\end{remark}
Proposition \ref{prop:1} states that for sufficiently large values of $L$, we may consider the new random sequence $\tilde{Z}^H$ as white noise. The speed of convergence of $K_{\tilde{Z}^H}(q)$ to the power spectrum of a white noise depends on the regularity of the initial power spectrum $S_{Z^H}(\omega)$. In \citep{LacazeRoviras2002} it is shown for NRZ and Biphase processes that their permuted sequences converge to a white noise for $L\geq500$. For the fGn, we can repeat the same experiment done in Figure \ref{fig:ACF1} starting from fBm with $H=\{0.25,0.75\}$ and $N = 4096$. Applying the random permutation vector $b=\left(b_0,b_1,\ldots,b_{L-1}\right)$, with the permutation length $L$ equal to the entire length $N-a$ of the fGn, we compute the autocorrelation function of the permuted sequences $\tilde{Z}^H_{t,a}$. In Figure \ref{fig:ACF_mixed} the autocorrelation of $\tilde{Z}^H_{t,a}$ is the same as a white noise, even for very small values of $L$ (see $a=4000$). Therefore, this experiment suggests that for a random permuted fGn, the convergence of its power spectrum to that of a white noise is very fast. 

\begin{remark}
    Random permutation acts only on the order of the elements of a sequence and not on their distribution. So the structure of self-similarity is maintained
    \begin{equation}\nonumber
        \{Z_{i,a}^{H_0}\} \overset{d}{=} \{a^{H_0}Z_{i,1}^{H_0}\} \;\;\;\mathbf{\Rightarrow}\;\;\;\{Z_{\overline{i}L+b_{\underline{i}}+\phi,a}^{H_0}\} \overset{d}{=} \{a^{H_0}Z_{\overline{i}L+b_{\underline{i}}+\phi,1}^{H_0}\}.
    \end{equation}
\end{remark}
The above technique removes the intradependence issue. To overcome also the interdependent problem,
given a sampled fBm of length $N$ and considered the two associated fGns  $Z_{t,1}^H$ and $Z_{t,a}^H$ of length $N-1$ and $N-a$, respectively, we extract from them two random subsequences of length $T\ll N-a$ and use them to build the empirical cumulative distribution functions in \eqref{eq:empirical_delta}. Thus, in addition to applying random permutation to eliminate the internal correlation of each sequence, we also eliminate the correlation between the two sequences while maintaining the self-similarity structure.

\section{Computational pipeline}\label{sec:new_method}
In the following, we summarize the steps of the novel method proposed in Section~\ref{sec:random_method}, which will be employed in the simulation study presented in Section~\ref{sec:simulation}.
\begin{enumerate}[leftmargin=*]
    \item Set the Hurst exponent $H_0\in(0,1]$ and simulate an fBm $B_{t}^{H_0}$ of length $N$
    \item Set a compact set of timescales $\pazccal{A}=\llbracket\underline{a},\overline{a}\rrbracket$ with $\underline{a}=1$ and extract two fGns $Z_{t,1}^{H_0}$ and $Z_{t,\overline{a}}^{H_0}$
    \item Generate a temporal random permutation and from $Z_{t,1}^{H_0}$ and $Z_{t,\overline{a}}^{H_0}$ the two subsequences $\Tilde{Z}_{t,1}^{H_0}$ and $\Tilde{Z}_{t,\overline{a}}^{H_0}$ are extracted with length $T\ll N-\overline{a}$;
    \item After selecting a dense set of values $H \in (0,1]$ for testing self-similarity, the KS test is performed for each $H$ to compare $\Tilde{Z}_{t,1}^{H_0}$ and $\overline{a}^{-H} \Tilde{Z}_{t,\overline{a}}^{H_0}$ and the corresponding KS statistic $\hat{\delta}_{\Tilde{Z}_t^{H}}(\Psi_H)$ is obtained
    \item The self-similarity parameter is estimated as    
    \begin{align}
\hat{H}_0&=\argmin\limits_{H\in(0,1]} \,\,\hat{\delta}_{\Tilde{Z}_t^{H_0}}
(\Psi_H)\nonumber\\
&=\argmin\limits_{H\in(0,1]} \sup\limits_{x\in\mathbb{R}}\left\vert \Phi_{\Tilde{Z}_{t,1}^{H_0}}(x)-\Phi_{\Tilde{Z}_{t,1}^{H_0}}\left(\overline{a}^{H-H_0}x\right)\right\vert \nonumber \\
& = \argmin\limits_{H\in(0,1]} \sup\limits_{x\in\mathbb{R}}\left\vert \frac{1}{T}\sum\limits_{i=0}^{T-1}\mathbbm{1}_{\Tilde{Z}_{i,1}^{H_0}\leq x}-\frac{1}{T}\sum\limits_{i=0}^{T-1}\mathbbm{1}_{\overline{a}^{-H}\Tilde{Z}_{i,\overline{a}}^{H_0}\leq x} \right\vert \label{eq:estimemp}.
\end{align}
\item The statistic $\hat{\delta}_{\Tilde{Z}_t^{\hat{H}_0}}(\Psi_H)$ is compared with the critical values of the standard KS distribution. Clearly, the proximity between the theoretical and empirical significance levels is studied for $\hat{H}_0=H_0$. This assumption is necessary to isolate the issue of the variance of estimator \eqref{eq:estimemp} from the intrinsic variability of the distribution of the KS statistic.
\end{enumerate}

\section{Simulation study}\label{sec:simulation}
In this section, we applied the method described in Section \ref{sec:new_method} to show that it solves both intradependent and interdependent problems, making the method introduced in 2004 more robust. In the next simulations, the length of fBm is set as $N=4096$. Any distribution is the result of $500$ simulations.

\subsection{Intradependent problem}
\begin{figure}[!ht]
    \centering
    \begin{subfigure}{1\textwidth}
        \centering
    \includegraphics[width=16cm,height=6.5cm]{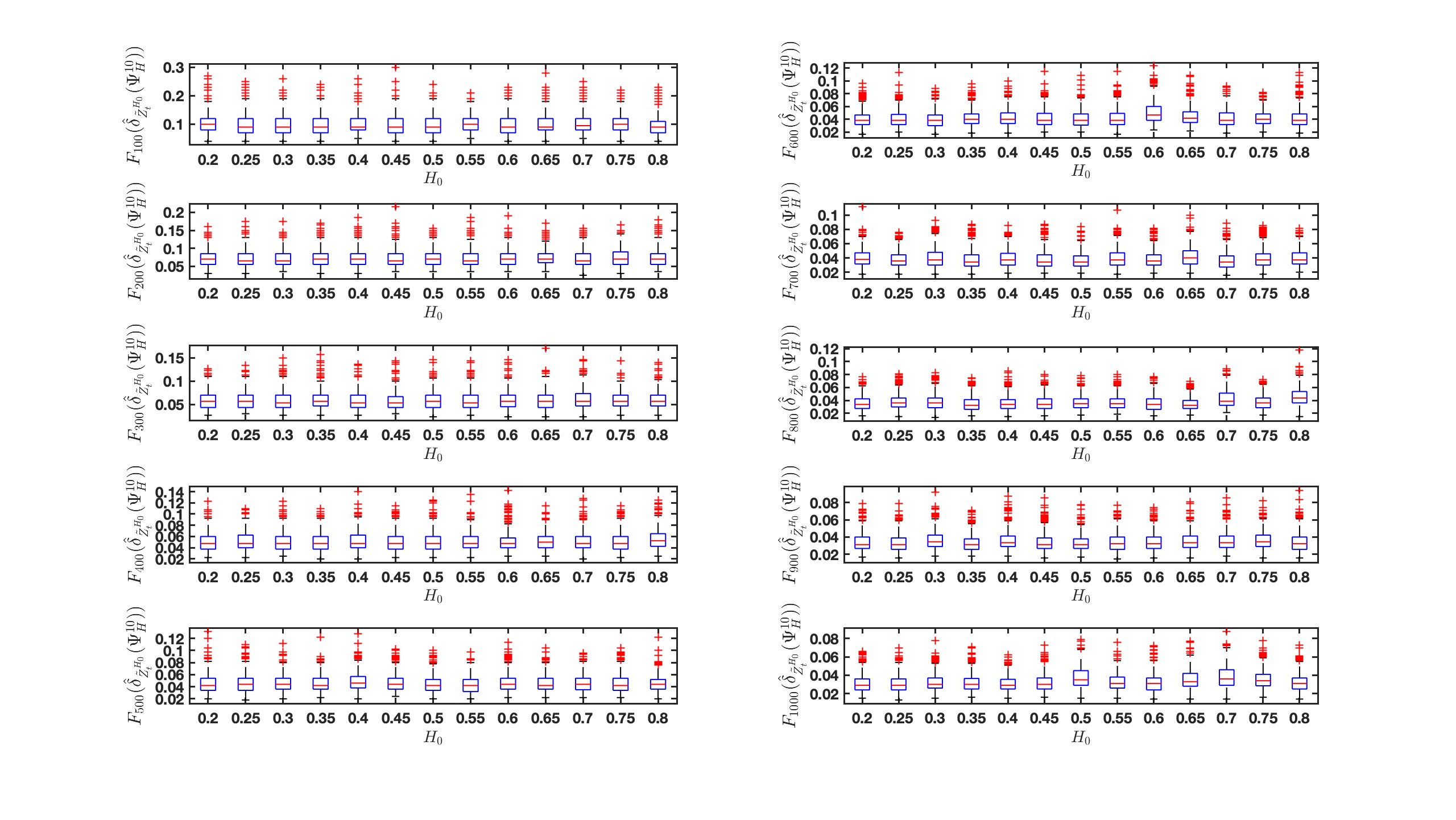}
        \caption{}
        \label{fig:delta VS H with a}
    \end{subfigure}
    \begin{subfigure}{1\textwidth}
        \centering
        \includegraphics[width=15.8cm,height=4.5cm]{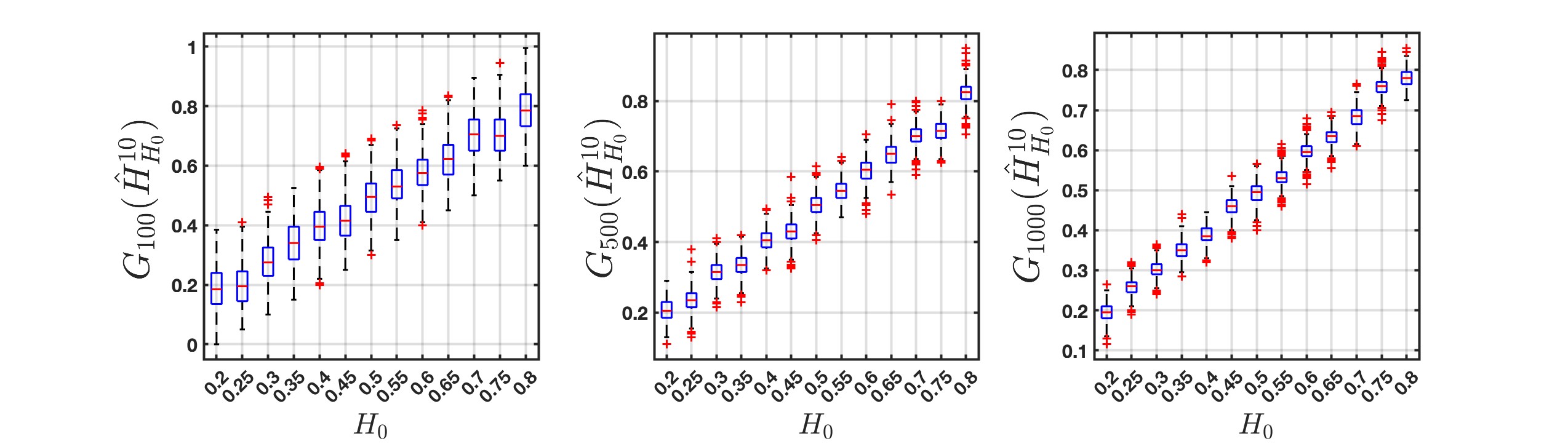}
        \caption{}
        \label{fig:H VS H with a}
    \end{subfigure}
    \caption{(a) Distributions $F_T\left(\hat{\delta}_{\tilde{Z}_t^{H_0}}(\Psi_H^{\overline{a}})\right)$ with $\overline{a}=10$, $T\in[100,1000]$ with $\Delta T=100$ and many self-similarity parameters $H_0 \in[0.2,0.8]$ with $\Delta H = 0.05$, remain stable as $H_0$ changes. (b) Analogous distributions are reported for $\hat{H}_{H_0}^10$ for $T=\{100,500,1000\}$. Each distribution is the result of $500$ simulations.}
\end{figure}
Step 1 in Section \ref{sec:new_method} was iterated for multiple values of $H_0 = [0.2,0.8]$ with $\Delta H = 0.05$ and setting $\overline{a} = 10$.

 We observe (Figure \ref{fig:delta VS H with a}) that the distributions $F_T\left( \hat{\delta}_{\Tilde{Z}_t^{H_0}}\right)$ remain stable -- i.e. no noticeable shifting biases are present in the plots -- with respect to $H_0$, with subsequence lengths $T\in[100,1000]$ and $\Delta T = 100$. The corresponding $G_T$ distributions are plotted in Figure $\ref{fig:H VS H with a}$ as $H_0$ varies. As we should expect, as $T$ increases, the numerosity of the data also increases and therefore the $\hat{H}$ estimate improves.
\subsection{Interdependent problem}
\begin{figure}[ht!]
    \centering
    \begin{subfigure}{0.45\textwidth}
        \centering
        \includegraphics[width=8.1cm,height=6.5cm]{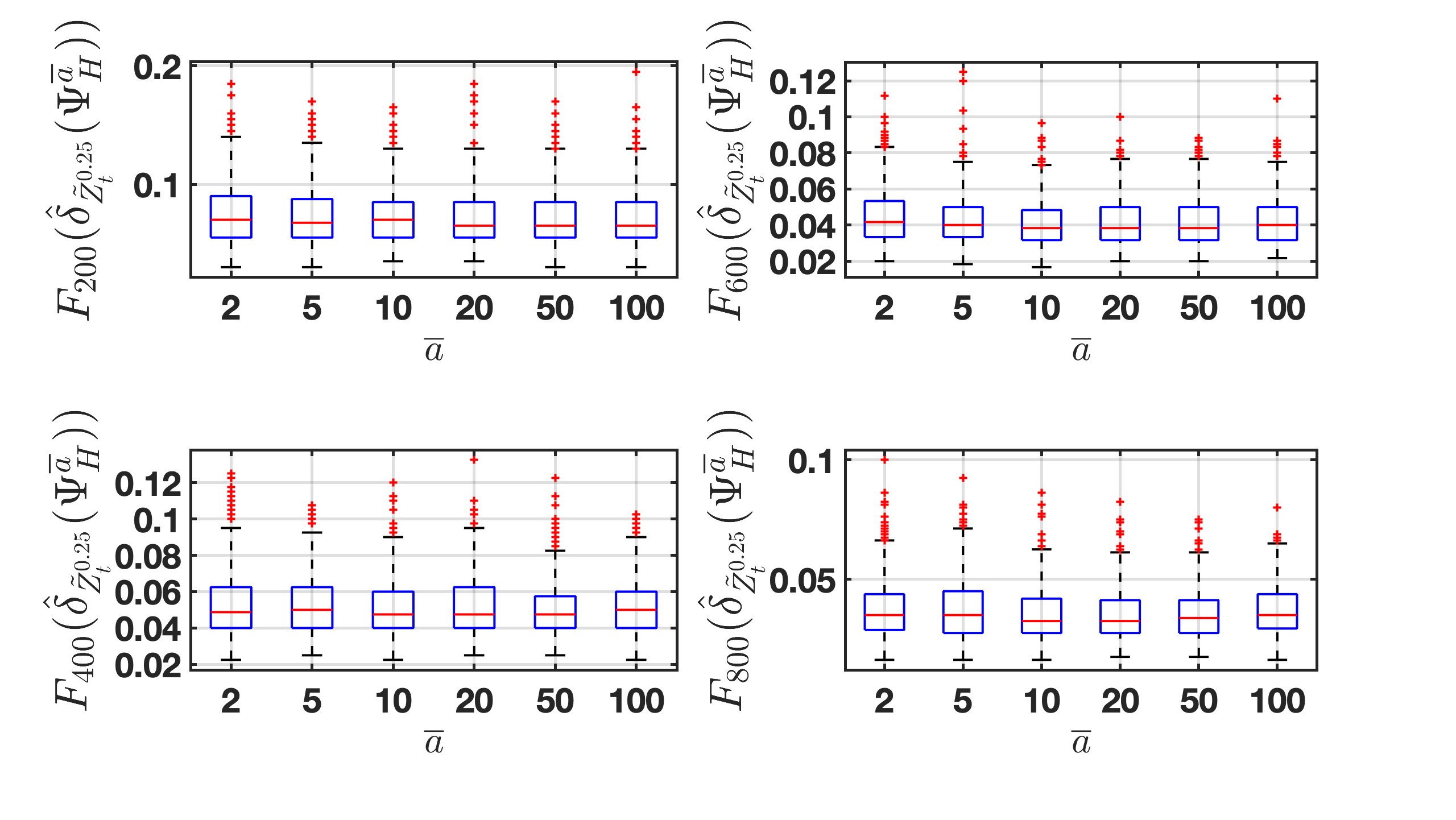}
        \caption{}
        \label{fig:H VS a 025}
    \end{subfigure}
    \hspace{0.1cm}
    \begin{subfigure}{0.45\textwidth}
        \centering
        \includegraphics[width=8.1cm,height=6.5cm]{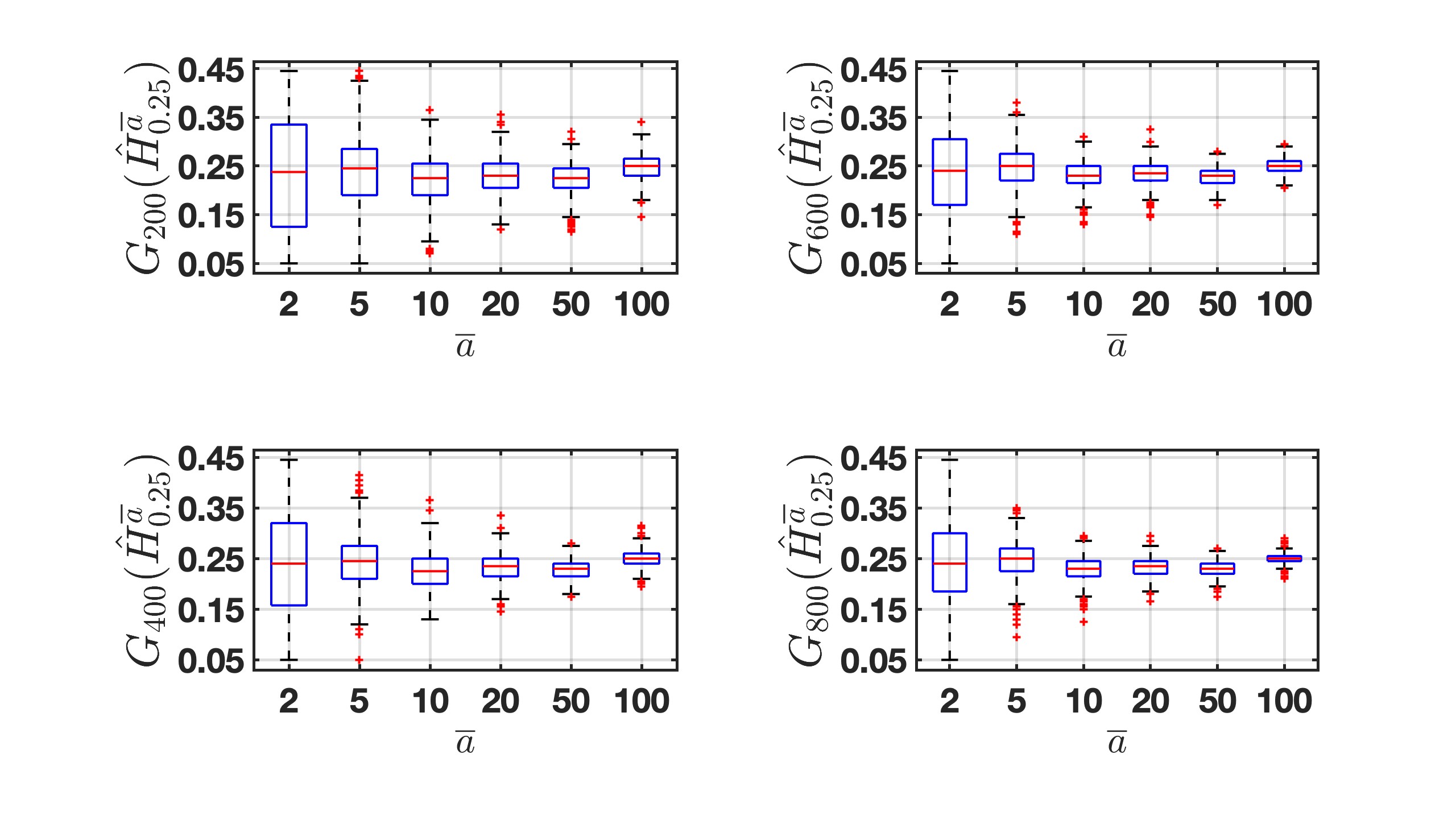}
        \caption{}
        \label{fig:H VS H 025}
    \end{subfigure}
    \begin{subfigure}{0.45\textwidth}
        \centering
        \includegraphics[width=8.1cm,height=6.5cm]{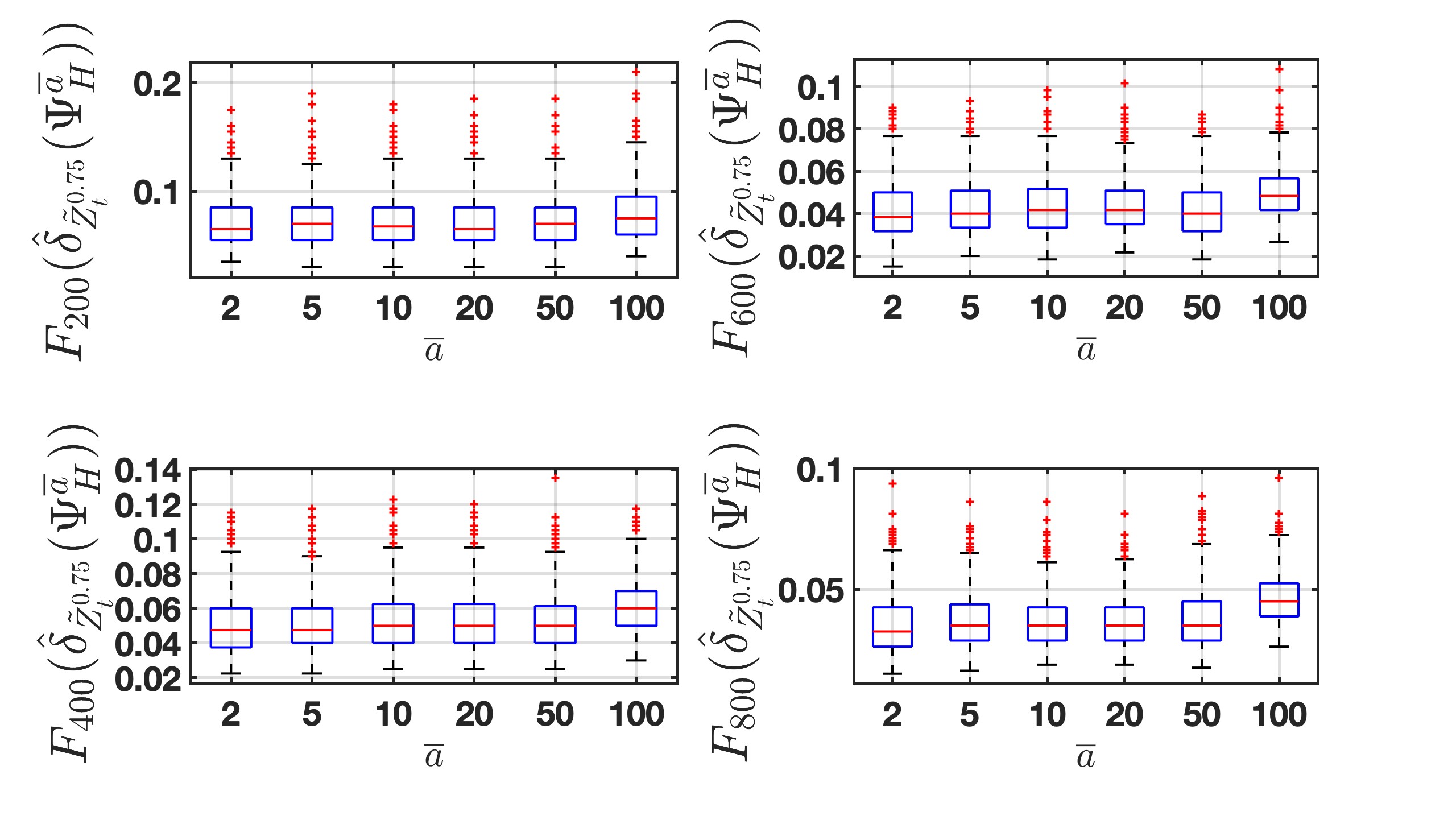}
        \caption{}
        \label{fig:H VS a 075}
    \end{subfigure}
    \hspace{0.1cm}
    \begin{subfigure}{0.45\textwidth}
        \centering
        \includegraphics[width=8.1cm,height=6.5cm]{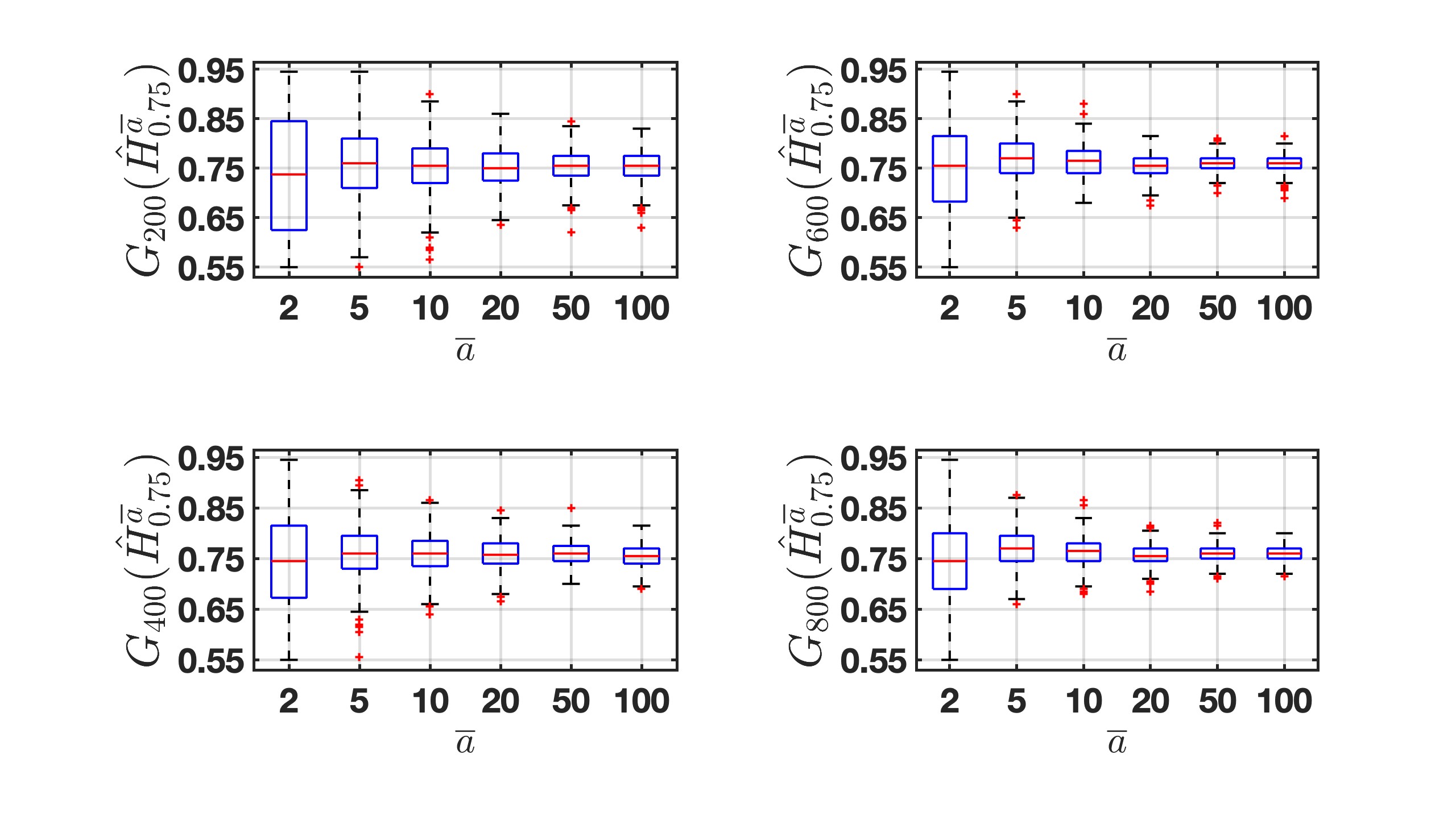}
        \caption{}
        \label{fig:H VS H 075}
    \end{subfigure}
    \caption{Study of the distributions $F_T(\hat{\delta}_{\tilde{Z}_t^{H_0}}(\Psi_H^{\overline{a}}))$ as the time scaling parameter changes for $H_0 = 0.25$ (panel (a)) and $H=0.75$ (panel (c)). Respectively shown are the distributions $G_T(\hat{H}_{H_0}^{\overline{a}})$ as $\overline{a}$ varies for $H=0.25$ (panel (b)) and $H=0.75$ (panel (d)). In these simulations $T=\{200,400,600,800\}$ and $\overline{a}=\{2,5,10,20,50,200\}$ were chosen. Each distribution is the result of $500$ simulations.}
\end{figure}

To show that the new method does not depend on the scaling factor, we iterated step 2 in section \ref{sec:new_method} for different values of $\overline{a}$ and for $H=0.25$ (Figure \ref{fig:H VS a 025}) and $H=0.75$ (Figure \ref{fig:H VS a 075}). The estimates of $H$ are shown in Figure \ref{fig:H VS H 025} and Figure \ref{fig:H VS H 075}, respectively.
By comparing these simulations with those without randomization in Figure~\ref{fig:delta VS a no mixed}, we observe that the distributions $F_T(\hat{\delta})$ exhibit greater stability, and the variances of the $G_T(\hat{H}_{H_0})$ distributions decrease as the length $T$ of the subsequence 
increases. Furthermore, while no noticeable growth is observed in the distributions for $H = 0.25$, the opposite trend is evident for $H = 0.75$ at high values of $\overline{a}$. Nonetheless, as shown in \citep{chakravartiLahaRoy1967}, the KS test remains effective even for sample sizes as small as $T \sim 50$. This ensures that the issue can be mitigated by maintaining low values of $T$.

\begin{table}[!ht]
    \centering
\caption{\label{tab:confidence_level} Empirical significance levels $\alpha$ for different values of $\overline{a}$ and Hurst exponent $H$ ($\underline{a}=1$ in all cases). Values were estimated from fBms of length $N=4096$ with subsequences of length $T=100$, using $M=1000$ simulations.}
    \renewcommand{\arraystretch}{1}
    \setlength{\tabcolsep}{8pt}
    \begin{tabular}{c cccc}
    \toprule
        \multicolumn{5}{c}{$\overline{a} = 10$} \\ 
        \midrule
        $H_0$ & $\alpha = 0.1$ & $\alpha = 0.05$ & $\alpha = 0.025$ & $\alpha = 0.01$ \\
        \midrule
        0.2 & 0.075 & 0.039 & 0.020 & 0.006 \\
        0.3 & 0.073 & 0.035 & 0.021 & 0.007 \\
        0.4 & 0.075 & 0.035 & 0.029 & 0.005 \\
        0.5 & 0.086 & 0.035 & 0.018 & 0.007 \\
        0.6 & 0.081 & 0.035 & 0.026 & 0.007 \\
        0.7 & 0.074 & 0.041 & 0.027 & 0.005 \\
        0.8 & 0.084 & 0.040 & 0.026 & 0.003 \\
    \midrule
        \multicolumn{5}{c}{$\overline{a} = 50$} \\
    \midrule
        0.2 & 0.075 & 0.030 & 0.024 & 0.007 \\
        0.3 & 0.083 & 0.039 & 0.029 & 0.010 \\
        0.4 & 0.101 & 0.035 & 0.035 & 0.006 \\
        0.5 & 0.084 & 0.037 & 0.027 & 0.006 \\
        0.6 & 0.106 & 0.054 & 0.032 & 0.005 \\
        0.7 & 0.105 & 0.036 & 0.030 & 0.007 \\
        0.8 & 0.107 & 0.046 & 0.022 & 0.007 \\
    \midrule
        \multicolumn{5}{c}{$\overline{a} = 100$} \\
    \midrule
        0.2 & 0.099 & 0.040 & 0.032 & 0.006 \\
        0.3 & 0.075 & 0.040 & 0.027 & 0.006 \\
        0.4 & 0.077 & 0.047 & 0.034 & 0.013 \\
        0.5 & 0.099 & 0.036 & 0.033 & 0.012 \\
        0.6 & 0.111 & 0.063 & 0.042 & 0.007 \\
        0.7 & 0.101 & 0.040 & 0.025 & 0.013 \\
        0.8 & 0.121 & 0.053 & 0.042 & 0.007 \\
    \bottomrule
    \end{tabular}
\end{table}

\section{Conclusion}\label{sec:conclusion}

This study revisits the Kolmogorov-Smirnov (KS) test applied to estimate the self-similarity parameter in fractional processes, highlighting its limitations in handling data dependencies. The inherent intradependencies and interdependencies within fractional processes undermine the reliability of traditional KS-based approaches. 

To overcome these challenges, we proposed a random permutation-based method that effectively disrupts the autocorrelation structure of fractional Gaussian noise while preserving the self-similarity property of the process. The slow decay of the autocorrelation function of fGn ensures that the method can be effectively applied to other fractional processes, whose autocorrelation functions decay at least as fast as, if not faster than, that of the fGn.
Simulation results validate the robustness of the proposed approach, exhibiting consistent performance across a range of Hurst exponent values and scaling factors. 

By enhancing the applicability of distribution-based self-similarity estimation, this method provides a reliable framework for analyzing fractional processes in diverse fields such as turbulence, finance, and telecommunications. Future research directions include extending this methodology to multidimensional or non-Gaussian processes, further broadening its scope and applicability.

\textbf{Declaration of generative AI in scientific writing} During the preparation of this work, the authors used ChatGPT in order to improve the readability and language of the manuscript. After using this tool/service, the authors reviewed and edited the content as needed and take full responsibility for the content of the published article.


\bibliographystyle{plain}
\bibliography{Bibliography} 
\end{document}